\documentclass[journal=jctcce,manuscript=article]{achemso}
\usepackage{graphicx} 
\usepackage{dcolumn}
\usepackage{bm}
\usepackage[utf8]{inputenc}
\usepackage{amssymb}
\usepackage{amsmath}
\usepackage{amsthm}
\usepackage{physics}
\usepackage{braket}
\usepackage{hyperref}
\setkeys{acs}{doi = true}
\usepackage{mathrsfs}
\usepackage{xcolor}
\usepackage{mathtools}
\usepackage[title]{appendix}
\usepackage[acronym,nonumberlist]{glossaries}
\usepackage{dsfont}
\usepackage{qcircuit}
\usepackage{textcomp}
\usepackage{orcidlink}
\usepackage[version=4]{mhchem}
\usepackage{amsfonts}
\usepackage{subcaption}
\usepackage{bbold}

\makenoidxglossaries

\newacronym{qksd}{QKSD}{quantum Krylov subspace diagonalization}
\newacronym{qpe}{QPE}{quantum phase estimation}
\newacronym{spe}{SPE}{statistical phase estimation}
\newacronym{qsvt}{QSVT}{quantum singular value transform}
\newacronym{qsp}{QSP}{quantum signal processing}
\newacronym{df}{DF}{double factorization}
\newacronym{thc}{THC}{tensor hypercontraction}
\newacronym{lcu}{LCU}{linear combination of unitaries}
\newacronym{bliss}{BLISS}{block-invariant symmetry-shift}
\newacronym{dmrg}{DMRG}{density matrix renormalization group}
\newacronym{mps}{MPS}{matrix product state}
\newacronym{casci}{CASCI}{complete active space configuration interaction}
\newacronym{cdf}{CDF}{cumulative distribution function}
\newacronym{rmse}{RMSE}{root mean squared error}

\newcommand{\lspan}{\mathrm{span}}
\newcommand{\e}{\mathrm{e}}

\title{Optimizing and Comparing Quantum Resources of Statistical Phase Estimation and Krylov Subspace Diagonalization}

\author{Oumarou Oumarou}
\affiliation{Covestro Deutschland AG, Leverkusen, Nordrhein-Westfalen 51373, Germany}
\email{oumarou.oumarou@covestro.com}

\author{Pauline J. Ollitrault}
\affiliation{QC Ware Corp, Palo Alto, California 94306, USA and Paris, Île-de-France 75003, France}
\email{pauline.ollitrault1@gmail.com}

\author{Stefano Polla}
\affiliation{Lorentz Institute for Theoretical physics, Leiden University, The Netherlands}
\email{polla@lorentz.leidenuniv.nl}

\author{Christian Gogolin}
\affiliation{Covestro Deutschland AG, Leverkusen, Nordrhein-Westfalen 51373, Germany}
\alsoaffiliation{Lorentz Institute for Theoretical physics, Leiden University, The Netherlands}
\email{christian.gogolin@covestro.com}

\date{September 2025}

\begin{document}

\begin{abstract}
    We develop a framework that enables direct and meaningful comparison of two early fault-tolerant methods for the computation of eigenenergies, namely \gls{qksd} and \gls{spe}, within which both methods use expectation values of Chebyshev polynomials of the Hamiltonian as input.
    For \gls{qksd} we propose methods for optimally distributing shots and ensuring sufficient non-linearity of states spanning the Krylov space.
    For \gls{spe} we improve rigorous error-bounds, achieving roughly a factor $2/3$ reduction of circuit depth.
    We provide insights into the scalability of and the practical realization of these methods by computing the maximum Chebyshev degree, linearly related to circuit depth, and the respective number of repetitions required for the simulation of molecules with active spaces up to 54 electrons in 36 orbitals by leveraging \gls{mps}/\gls{dmrg}.
\end{abstract}

\maketitle


\section{Introduction}
\label{sec:introduction}

The first generations of error-corrected quantum computers are expected to operate at relatively low code distances and will therefore remain limited in circuit depth capacity, making it essential to develop algorithms with intrinsic resilience to residual noise.
At the same time, the slower logical gate times, compared to the physical clock speed of the qubits, will impose even stricter requirements on the maximum number of repetitions that are acceptable.
Owing to their favorable depth–sampling tradeoffs, \Gls{qksd}~\cite{Kirby_2023} and \gls{spe}~\cite{Wan_2022, Lin_2022} have consequently been proposed as promising approaches for simulating quantum many-body systems, such as molecular Hamiltonians, on early fault-tolerant quantum computers.

Both algorithms rely on similar quantum primitives but extract ground-state energies in fundamentally different ways. At high level, \gls{qksd} builds a low-dimensional approximation of the Hamiltonian by generating a Krylov subspace from an initial state using polynomial functions of the Hamiltonian, and then classically solves a projected generalized eigenvalue problem within this subspace. Its accuracy improves as the dimension of the Krylov space (and thus the maximal polynomial degree) increases. In contrast, \gls{spe} reconstructs the cumulative spectral distribution associated with the initial state by approximating a Heaviside function through a truncated polynomial expansion, and then locates the ground-state energy via a classical search procedure. In the formulation adopted here, both methods use expectation values of Chebyshev polynomials implemented through qubitized walk operators~\cite{low2019hamiltonian}, enabling a direct and meaningful comparison of their circuit depth and sampling requirements.

In this work, we improve performance guarantees and establish advantageous strategies for the practical application of \gls{spe} and \gls{qksd}.
We estimate the number of required repetitions $M$ and maximum Chebyshev polynomial degree $K$ (proportional to circuit depth and non-Clifford gate count) to compute chemically precise ground state energies of molecular systems with active space sizes approaching the boundary of what can still be classically addressed with \gls{dmrg}.
We use active space Hamiltonians compressed and optimized with a combination of \gls{thc}~\cite{lee2021even} and \gls{bliss}~\cite{loaiza2023block, patel2025global, caesura2025faster}, corresponding to block encoding normalization factors $\lambda_{\mathrm{THC-BLISS}}$ and spectral shift $\beta_{\mathrm{THC-BLISS}}$. 

We perform the resource estimates in a setting where both methods use the quantum computer to evaluate expectation values of Chebyshev polynomials of the Hamiltonian, realized with qubitized walk operators.
In this formulation, \gls{spe} actually achieves Heisenberg scaling, in the sense that the maximal polynomial degree $K$, and hence the circuit depth, scales linearly with the inverse target precision, $K \in \mathcal{O}(\delta^{-1})$. Furthremore, in this setting, \gls{spe} estimates the arc-cossine of the ground state energy and therefore can potentially benefits from the error propagation, when inverting the arc-cosine, similar to that exhibited in \gls{qpe} \citenum{Poulin_2018,Babbush_2018} which can be used to amplify the lower part of the spectrum. We find however that for the \gls{thc}-\gls{bliss} Hamiltonians of the considered molecules the effect is negligible (see Section~\ref{sec:res_spe_benchmarks} for details). 
The number of samples required to estimate the approximate \gls{cdf} contributes an additional polylogarithmic dependence on $\delta^{-1}$, resulting in a total number of samples that depends modestly with respect to $\delta$. 

In \gls{qksd}, the use of Chebyshev polynomials leads to projection matrices that can be written as linear combinations of Toeplitz- and Hankel-structured matrices. Consequently, only $2K$ expectation values are required to construct the $K\times K$ projection matrices.
However, the scaling behavior of $K$ differs from that of \gls{spe}. In principle, $K$ scales only logarithmically with the inverse precision, whereas the number of repetitions goes as $\delta^{-2}$~\cite{Kirby_2023}. It should be noted, however, that existing circuit depth bounds are not tight, and as we demonstrate in subsequent sections, $M$ and $K$ are correlated. 
For these reasons, the resource estimates for \gls{qksd} presented in this work are not derived from asymptotic scaling bounds but rather obtained by classical emulation of the algorithm. 
Concretely, for a given $K$, we first determine the systematic error that persists even in the infinite-shot limit. We then optimize the shot allocation strategy to find the minimal total shot count $M$ required to reduce the shot-noise-induced standard error below $10^{-3}$~Ha.
 
To enable a meaningful comparison between the two methods, we use the improved error bounds for \gls{spe} derived in Section~\ref{sec:res_spe} to identify the circuit depth and number of samples required for \gls{spe} to achieve an error less than or equal to the root-mean-square error of \gls{qksd}, which incorporates both systematic and statistical contributions.

\begin{figure}[h]
    \centering
    \includegraphics[width=0.7\linewidth]{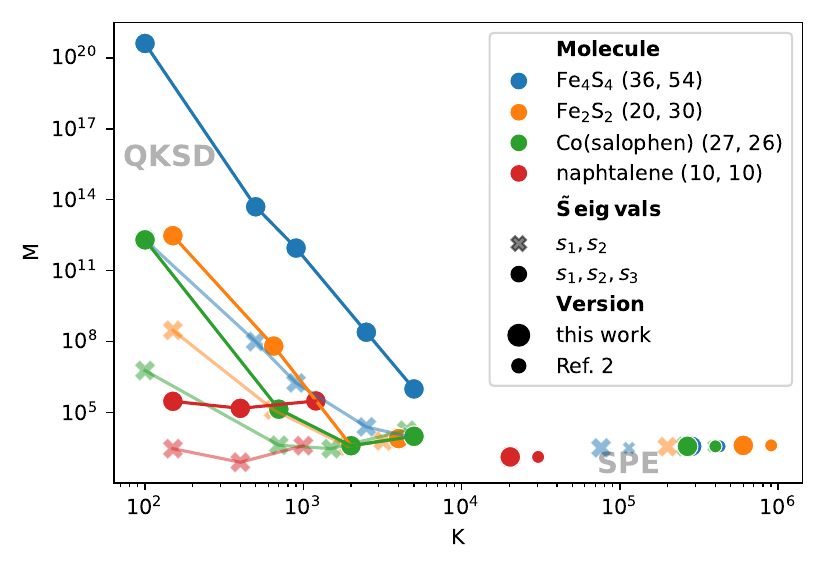}
    \caption{
Maximum Chebyshev polynomial degree $K$ (proportional to circuit depth) and total shot count $M$ required to achieve a shot-noise error below $10^{-3}$~Hartree for various molecules. For \gls{qksd} (symbols connected by lines), the generalized eigenvalue problem is solved while retaining the dominant two (crosses) or three (circles) eigenvalues of the overlap matrix $\widetilde{S}$ (see Table~\ref{tab:data} for the achieved \acrfull{rmse} and additional details).
The scattered points indicate the circuit depth and shot count for which, with 99\% success probability, the \gls{spe} is guaranteed to achieve an error of $10^{-3}$~Hartree or be on par with the \gls{rmse} of \gls{qksd} at the largest $K$ shown if that value is higher.
Small markers correspond to $M$ and $K$ estimates with the error bounds from Ref.~\citenum{Wan_2022}, while large markers are based on improved bounds derived in Section~\ref{sec:tight_bounds} (see Figure~\ref{fig:delta_E_vs_K} for an analysis of their tightness).
}
    \label{fig:fig1}
\end{figure}

Figure~\ref{fig:fig1} summarizes the resulting resource estimates, providing a concise overview of the circuit depth and sampling requirements before we present the detailed methodology.
As this figure illustrates, the total number of samples for \gls{qksd} significantly decreases as the Krylov subspace size grows, up to the point where it becomes comparable to the total number of observables to be measured.
This is particularly evident for smaller molecules such as naphthalene.

In contrast, even with the improved error bound found in this work, \gls{spe}, while consistently requiring $M \leq 10^{5}$ shots, does need much larger maximum polynomial degrees $K$, and hence deeper circuits, than \gls{qksd}.
Since \gls{spe} determines its energy estimate via binary search, one can, occasionally get an energy very close to the true ground state energy, making the comparison subtle.
We therefore report the $K$ at which the accuracy of \gls{spe}, in case of a successful binary search, is guaranteed to be on par with that of the mean squared error (combining the systematic and shot noise error) of \gls{qksd}.
The success probability $p_{success}$ of the binary search can be increased by multiplying the reported shot counts $M$ with a logarithmic prefactor $\log((1-p_{success})^{-1})$.

In summary, we find that by optimally distributing shots and choosing a large enough polynomial degree, \gls{qksd} seems to be able to achieve the same precision as \gls{spe} for a comparable numbers of total shots $M$, but about one order of magnitude lower polynomial degree $K$.
If $K$ is further reduced, the number of shots $M$ required to maintain the same level of accuracy increases steeply. 

We find that the number of samples required for \gls{qksd} is largely dominated by the need to reduce the noise on the eigenvalues of the overlap matrix $\widetilde{S}$ in the generalized eigenvalue problem (see Section~\ref{sec:qksd} for details).
This noise must be sufficiently low to clearly distinguish the physically relevant eigenvalues from the rest of the noisy spectrum.
For the systems studied here with \gls{qksd}, achieving chemical accuracy is generally accomplished by restricting the generalized eigenvalue problem to the subspace defined by the three leading eigenvectors of the overlap matrix.
As the Krylov dimension $K$ increases, these three leading eigenvalues of $\widetilde{S}$ grow and their separation from the rest of the spectrum widens. This increasing spectral gap is the primary reason for the sharp reduction in the number of shots required at larger $K$ (see Section~\ref{sec:shot_noise_numerics}).

Evidently, the overlap of the initial state affects both $K$ and $M$.
In \gls{spe} the effect of $p_0$ on $K$ is relatively marginal thanks to the logarithmic scaling, in contrast to the total number of samples $M$ which asymptotically scale as the squared inverse of $\mathcal{O}(p_0^{-2})$.
In \gls{qksd} the asymptotic scalings provided in Ref.~\citenum{Kirby_2023} narrates a qualitatively similar story where $K$ asymptotically scales logarithmically with $p_0$, whereas $M$ possesses a quartic scaling $\mathcal{O}(p_0^{-4})$.
For lower $p_0$ than assumed here, one can thus expect the $M$ needed for \gls{qksd} to rise faster than those for \gls{spe}.

For reference, we also provide a back-of-the-envelope calculation estimating the resources that would be required to recover the same results using a quantum phase estimation with a multi-qubit control register.
In order to achieve Heisenberg limit, we consider the QPE variant using a sine probe-state \cite{babbushEncoding2018}. 
A single sample of this QPE circuit using $K$ calls to the walk operator achieves Holevo error $\tan(\frac{\pi}{K+1})\approx\frac{\pi}{K}$ on the sampled eigenphase of the qubitized walk operator. 
As long as this error is much smaller than the separation between one eigenphases and the others, in the noiseless setting one can distinguish samples originated by that eigenphase.
Repeating sampling of the circuit $M$ times, we expect an average of $M p_0$ samples coming from the ground-state eigenphase (where $p_0$ is the squared overlap of the prepared intial state and ground state). 
Combining these samples through maximum-likelihood estimation \cite{Dutkiewicz_2025} (a simple average would result in estimator bias), one can achieve an expected error of approximately $\frac{\pi}{K \sqrt{M p_0}}$.

\begin{table}[]
    \centering
    \begin{tabular}{l|r|r|r|r}
         & $\mathrm{Fe_4S_4} (54, 36)$ & $\mathrm{Fe_2S_2} (30, 20)$ & Co(salophen) (26, 27) & Naphthalene (10, 10) \\
         \hline
     $\Delta E_0$ & 0.014105 &  0.038936 & 0.061914 & 0.093205  \\    
     \hline 
     \gls{rmse} $s_1, s_2$ & 0.003413 & 0.000501 & 0.000444 & 0.001 \\
     \hline 
     \gls{rmse} $s_1, s_2, s_3$ & 0.000914 & 0.000166 & 0.000431 & 0.001  \\
     \hline 
     $\lambda_{\mathrm{THC-BLISS}}$ & 63.355 & 24.390 & 28.132 & 5.492 \\
     \hline
     $\beta_{\mathrm{THC-BLISS}}$ & 336.353 & 118.880 & 2402.159 & 382.282 \\
     \hline
     $p_0$ & 0.5 & 0.5 & 0.5 & 0.710886   \\
    \end{tabular}
    \caption{
    Summary of system parameters and \gls{qksd} performance for the molecular active spaces considered in this work. $\Delta E_0$ is the energy difference between the initial state and the ground state (in Hartree). \gls{rmse} of \gls{qksd} (in Hartree) when truncating the generalized eigenvalue problem to the two ($s_1, s_2$) or three ($s_1, s_2, s_3$) largest eigenvalues of the overlap matrix $\widetilde{S}$, evaluated at the largest $K$ for which data is shown in Figure~\ref{fig:fig1}. For the naphtalene case, we fixed the \gls{rmse} to 1 mH as a target accuracy for \gls{spe}), as the actual achieved \gls{rmse} is too small. Also listed are the THC-BLISS norm $\lambda_{\mathrm{THC-BLISS}}$ and shift $\beta_{\mathrm{THC-BLISS}}$, as well as the initial state overlap $p_0$ (see Section~\ref{sec:molecules} for details on the initial state construction).
    Active space Hamiltonians are taken from Ref.~\citenum{li2017spin} for the iron-sulfur complexes and from Ref.~\citenum{Dutkiewicz_2025} for Co(salophen) and naphthalene.
    }
    \label{tab:data}
\end{table}

\begin{table}[]
    \centering
    \begin{tabular}{l|r|r|r|r}
        & $\mathrm{Fe_4S_4} (54, 36)$ & $\mathrm{Fe_2S_2} (30, 20)$ & Co(salophen) (26, 27) & Naphthalene (10, 10) \\
        \hline
        $\delta$ $s_1, s_2$  & 0.001368 & 0.000433 & 0.000207 & 0.000416\\
        \hline
        $\delta$ $s_1, s_2, s_3$ & 0.000390 & 0.000103 & 0.000216 & 0.000416\\
    \end{tabular}
    \caption{Ground state energy error $\delta$ (in Hartree) achieved by \gls{spe} corresponding to the data reported in Figure~\ref{fig:fig1}.
    In some cases the achieved error is roughly a factor of two to three better than promised by the bound used to pick $M$ and $K$ in Figure~\ref{fig:fig1}, however in other cases the bound is nearly tight (see Figure~\ref{fig:delta_E_vs_K} for a detailed analysis) 
    } 
    \label{tab:placeholder}
\end{table}
\section*{Outline}
The remainder of the paper is organized as follows. Section~\ref{sec:theory} provides a concise overview of both \gls{qksd} and \gls{spe}. Section~\ref{sec:tight_bounds} establishes improved bounds on the Heaviside truncation error, yielding tighter resource estimates in terms of circuit depth and sample count. Section~\ref{sec:acdf_cheby} derives an approximation of the \gls{cdf} using expectation values of Chebyshev polynomials of the Hamiltonian, which invoke qubitized walk operators in place of time evolution. Section~\ref{sec:qksd_numerics} examines the effect of the energy window size, Hamiltonian shift and normalization, and the subsampling step size on \gls{qksd} performance. Section~\ref{sec:shot_noise_numerics} assesses the validity of the proposed shot-allocation heuristics in the presence of sampling noise. Finally, Section~\ref{sec:conclusion} summarizes the main findings and discusses their implications.

\section{Theory}
\label{sec:theory}
This section provides an overview of the \gls{qksd} and \gls{spe} methods, focusing on the specific formulations employed in this work.
%
We begin by introducing notation.
Given a physical Hamiltonian $\hat{\mathcal{H}}$, such as a molecular electronic Hamiltonian, the quantum algorithms considered here will operate on a shifted $\hat{H}' \coloneqq \hat{\mathcal{H}} - \beta \hat{I}$ and/or shifted and rescaled $\hat{H}\coloneqq\frac{\hat{H}'}{\lambda}$ version of $\hat{\mathcal{H}}$.
The scalar shift $\beta$ and normalization factor $\lambda$ need to be chosen such that $\hat{H}$ is unit less and $\|\hat{H}\|\leq 1$ (in Table~\ref{tab:norms_shifts} we compare different methods to find suitable $\beta$ and $\lambda$).
We then denote by $\hat{U}$ the block encoding of $\hat{H}$ defined as
\begin{align}
    \bra{0}_a\hat{U}\ket{0}_a\ket{\phi}_s = \frac{\hat{H}'}{\lambda}\ket{\phi}_s \coloneqq \hat{H}\ket{\phi}_s.
\end{align}
The subscripts $a$ and $s$ identify the ancillary register and system qubits of the block encoding, respectively, and the state $\ket{0}_a$ flags the subspace of the system qubits.
We will omit these subscripts for better readability, in the cases where they can be inferred from context.

With $\hat{R}\coloneqq\left( 2\ket{0}_a\bra{0}_a-I_a \right) \otimes I_s$, the qubitized walk operator $\hat{\mathcal{W}}$ is then defined as
\begin{align}
    \hat{\mathcal{W}}\coloneqq \hat{R}\hat{U}.
\end{align}

We further denote by $\{\lambda_k\}_{k=0}^{N-1}$ and $\{\ket{\lambda_k}\}_{k=0}^{N-1}$ the lists of eigenvalues and corresponding eigenvectors of $\hat{H}$ so that the spectral decomposition of $\hat{H}$ can be written as
\begin{align}
    \hat{H} = \sum_{i=0}^{N-1}\lambda_i\ket{\lambda_i}\bra{\lambda_i}.
\end{align}
The physical eigenenergies $\{E_k\}_{k=0}^{N-1}$ of $\hat{\mathcal{H}}$ are related to $\{\lambda_k\}_{k=0}^{N-1}$ through simple rescaling and shifting
\begin{align}
    E_k = \beta + \lambda \cdot \lambda_k.
\end{align}

\subsection{Krylov Subspace Diagonalization}
\label{sec:qksd}
The \gls{qksd} method approximates the eigenspectrum of $\hat{H}$ with that of its projection $\widetilde{H}$ onto a so-called Krylov subspace $\mathcal{K}$ spanned by $K+1$ Krylov states $\ket{\psi_k}$.
The Krylov states are generated by applying a function $f_k(\hat{H})$, depending on and commuting with the Hamiltonian $\hat{H}$, to a fixed initial state $\ket{\psi_0}$
\begin{align}
    \label{eq:krylov}
    \ket{\psi_k} &\coloneqq f_k(\hat{H})\ket{\psi_0}, \\
    \mathcal{K} &\coloneqq \lspan\big(\{\ket{\psi_k}\}_{k=0}^{K}\big) .
\end{align}
 Various classes of functions $f$ have been considered in the literature, including time evolution with real or imaginary time \cite{Cohn_2021, piccinelli2025quantum, Motta_2019}
and polynomials\cite{Kirby_2023}.
Specifically Chebyshev polynomials of the Hamiltonian, possess certain advantageous properties, such as favorable shot budget \cite{kanasugi2025mirror}, particularly simple circuits \cite{Dong_2021} and a Toeplitz structure of the projected matrices\cite{Kirby_2023}.

Approximations $\tilde \lambda_m$ to at most $K+1$ of the true eigenvalues $\lambda_m$ of $\hat{\mathcal{H}}$, corresponding to eigenstates with which the initial state $\ket{\psi_0}$ has non-vanishing overlap, can then be obtained by solving the generalized eigenvalue problem
\begin{equation}
    \widetilde{H}\alpha_m = \tilde{\lambda}_m \widetilde{S} \alpha_m,
    \label{eq:gen_eigen}
\end{equation}
with
\begin{align}
    \widetilde{H}_{kj} &\coloneqq \bra{\psi_k}\hat{H}\ket{\psi_j}, \\
    \widetilde{S}_{kj} &\coloneqq \braket{\psi_k|\psi_j}.
\end{align} 
The principal appeal of \gls{qksd} is the exponential decay of the difference between the true eigenvalues $\lambda_m$ of $\hat{H}$ and the generalized Krylov eigenvalues $\tilde{\lambda}_m$ of $\widetilde{H}$ with the Krylov subspace size $K+1$, at least in the limit of a vanishing shot-noise.
In practice, however, the finite number of shots used to estimate $\widetilde{H}$ and $\widetilde{S}$ as well as the regularization of the overlap matrix add additional terms to the original error\cite{Kirby_2023}.
Nevertheless, the prospect of maintaining shallow circuits remains viable, provided that a sufficient shot budget is allocated to adequately suppress both statistical and regularization errors.

In the remainder of this work we exclusively consider Chebyshev polynomials $T_k$ of degree $k$ up to a maximum degree $K$ to generate the Krylov states $|\psi_k\rangle$ whenever talking about \gls{qksd}, i.e., we will define the Krylov subspace as
\begin{align}
    \mathcal{K}_{\mathrm{cheb}} \coloneqq \lspan\big(\{T_{k}(\hat{H}) \ket{\psi_0}\}_{k=0}^{K}\big). 
\end{align}
Consequently, we have
\begin{equation}
    \widetilde{H}_{kj}
    = \bra{\psi_0}\, T_{k}(\hat{H})\, \hat{H}\,  T_{j}(\hat{H})\ket{\psi_0}.
    \label{eq:norm_cheb}
\end{equation}
The Chebyshev polynomials $T_k$ obey the following relation
\begin{equation}
  T_k(\hat{H}) T_j(\hat{H}) = \tfrac{1}{2} \Big( T_{k+j}(\hat{H}) + T_{|k-j|}(\hat{H}) \Big)
\end{equation}
and 
\begin{equation}
  T_1(\hat{H}) = \hat{H}, \quad T_0(\hat{H}) = 1 . 
\end{equation}
For ease of notation, we denote $\langle T_j(\hat{H})\rangle \coloneqq \bra{\psi_0}T_j(\hat{H})\ket{\psi_0}$.
We therefore have
\begin{equation}
\begin{split}
\widetilde{H}_{kj} = \tfrac{1}{4} \Big( 
&\langle T_{k+j+1}(\hat{H}) \rangle 
+ \langle T_{|k+j-1|}(\hat{H}) \rangle \\
&+ \langle T_{|k-j+ 1|}(\hat{H}) \rangle 
+ \langle T_{|k-j- 1|}(\hat{H}) \rangle 
\Big)
\end{split}
\label{eq:qksd_H}
\end{equation}
and
\begin{equation}
\widetilde{S}_{kj} = \tfrac{1}{2} \Big( \langle T_{k+j}(\hat{H}) \rangle + \langle T_{|k-j|}(\hat{H}) \rangle \Big).
\label{eq:qksd_S}
\end{equation}
The $(K+1)\times(K+1)$ entries of $\widetilde{S}_{kj}$ and $\widetilde{H}_{kj}$ respectively can be determined from just $2K+2$ expectation values with respect to Chebyshev polynomials $\{\langle T_k(\hat{H})\rangle\}_{k=0}^{2K+1}$.
In fact, only $2K-1$ need to be determined via measurements under the assumption that $\bra{\psi_0} \hat{H} \ket{\psi_0}$ is known.

Polynomial functions of $\hat{\mathcal{H}}$ can be realized on a quantum computer by means of the \gls{qsp} where the special case of Chebyshev polynomials yields trivial angle parameters\cite{Dong_2021}.
Specifically, the block encoding of the Chebyshev polynomial of degree $k$ can be realized using the $k$-th power of the the walk operator $\hat{\mathcal{W}}$
\begin{align}
    T_k(\hat{H})\ket{\psi_0}_s = \bra{0}_a\hat{\mathcal{W}}^k\ket{0}_a\ket{\psi_0}_s
\end{align}
and therefore 
\begin{align}
    \bra{\psi_0}T_k(\hat{H})\ket{\psi_0} = \bra{0}_a\bra{\psi_0}_s\hat{\mathcal{W}}^k\ket{0}_a\ket{\psi_0}_s .
\end{align}
As shown in Ref.~\citenum{Kirby_2023}, these expectation values can be obtained by preparing the state $\ket{\psi}_{\lfloor k/2\rfloor}\coloneqq(\hat{R}\hat{U})^{\lfloor k/2\rfloor}\ket{\psi_0}$ and measuring $\hat{U}$ or $\hat{R}$ depending on the parity of $k$ as follows:
\begin{align}
\label{eq:<T>}
    \bra{\psi_0}T_k(\hat{H})\ket{\psi_0} = \begin{cases}
        \bra{\psi}_{\lfloor k/2\rfloor} \hat{U} \ket{\psi}_{\lfloor k/2\rfloor} & \text{if } k \text{ is odd} \\
        \bra{\psi}_{\lfloor k/2\rfloor} \hat{R} \ket{\psi}_{\lfloor k/2\rfloor} & \text{if } k \text{ is even}
    \end{cases}
\end{align}

\subsection{Statistical Phase Estimation}
\label{sec:spe}
Let us begin by expressing the initial state $\ket{\psi_0}$ in an eigenbasis of $\hat{H}$
\begin{align}
    \ket{\psi_0} &= \sum_{i=0}^{N-1}\braket{\lambda_i| \psi_0}\ket{\lambda_i}, \text{ where } |\braket{\lambda_i| \psi_0}|^2\eqqcolon p_i.
\end{align}
We denote the cumulative and probability distribution functions by $C(x)$ and $p(x)$:
\begin{align}
    p(x) &\coloneqq\sum_{i=0}^{N-1}p_i\,\delta(x-\lambda_i)   \\
    C(x) &\coloneqq\sum_{i|\lambda_i\leq x}p_i = (\Theta \ast p)(x).    
\end{align}
where $\Theta(x)$ is the Heaviside function. 
Approximating the ground-state energy to $\delta$ accuracy is equivalent to finding a value $x^\star$ such that
\begin{align}
C(x^\star + \delta) > \eta / 2
\quad \text{and} \quad
C(x^\star - \delta) < \eta .
\label{eq:spe_gs_condition}
\end{align}
where $\eta$ is a lower bound estimate of $p_0$. 

In practice, the \gls{cdf} must be approximated. To this end, we start by approximating $\Theta(x)$, using the truncated expansion of the scaled error function introduced in Ref.~\citenum{Wan_2022}. The approximate Heaviside function reads
\begin{align}
    H(x) &\coloneqq F_0 + \sum_{j=0}^{K}F_{2j+1}\left(\e^{i(2j+1)x} - \e^{-i(2j+1)x}\right)
    \label{eq:Heavside_truncated}
\end{align}
where $K$ is a parameter controling the truncation error of the scaled error function and $F_k \in \mathbb{C}$ (see Ref.~\citenum{Wan_2022} for the expression of $F_k$). The corresponding approximate \gls{cdf},
\begin{align}
    \tilde{C}(x) &\coloneqq \left(H \ast p \right)(x) = F_0 + \sum_{j=0}^{K}F_{2j+1}\e^{i(2j+1)x}\left(\langle \e^{i(2j+1)\hat{H}}\rangle - \langle \e^{-i(2j+1)\hat{H}} \rangle\right), 
    \label{eq:acdf_estimator}
\end{align}
with $\langle\e^{-i(2j+1)\hat{H}}\rangle\coloneqq \bra{\psi_0} \e^{-i(2j+1)\hat{H}}\ket{\psi_0}$, satifies
\begin{equation}
    C(x-\delta)-\epsilon \leq \tilde{C}(x) \leq C(x+\delta)+\epsilon
    \label{eq:condition_acdf_ok}
\end{equation}
for all $x \in [-\frac{\pi-\delta}{2}, \frac{\pi-\delta}{2}]$ when $K=\mathcal{O}(\delta^{-1}\log(\epsilon^{-1}))$.

The conditions of Eq.~\ref{eq:spe_gs_condition} can then be inferred from $\tilde{C}$, by fixing $\epsilon=\eta/8$. In this case if $\tilde{C}(x)>5/8\eta$ then $C(x+\delta) > \eta/2$ and if $\tilde{C}(x)<7/8\eta$ then $C(x-\delta) < \eta$. 

In practice, we do not have access to the noiseless quantity $\tilde{C}$, but only to a finite-sampling estimate $\bar{G}$. If $\bar{G}(x) > \tfrac{3}{4}\eta$, we infer that the underlying noiseless value $\tilde{C}(x)$ is likely larger than $\tfrac{5}{8}\eta$, and therefore that $C(x+\delta) > \eta$. Conversely, if $\bar{G}(x) \le \tfrac{3}{4}\eta$, we infer that $\tilde{C}(x)$ is likely smaller than $\tfrac{7}{8}\eta$, which implies $C(x-\delta) < \eta$.

As in Ref.~\citenum{Lin_2022}, $x^\star$ is determined using a binary search procedure.
First, $x_{\mathrm{left}}$ and $x_{\mathrm{right}}$ are initialized to the extremes of the search interval. 
Then, at each iteration the value of $\bar{G}$ at midpoint $x_m\coloneqq \frac{x_{\mathrm{left}}+x_{\mathrm{right}}}{2}$ is evaluated.
If $\bar{G}(x_m)>\frac{3}{4}\eta$ then the left endpoint is moved to the right and updated to $x_{\mathrm{left}}=x_m+\frac{2}{3}\delta$.
Otherwise, the right endpoint is moved to the left and updated to $x_{\mathrm{right}}=x_m-\frac{2}{3}\delta$.
The procedure stops when $|x_{\mathrm{right}}-x_{\mathrm{left}}|\leq 2\delta$ and returns $x_m$ as $x^\star$ with a maximum error of $\delta$.

The variance of the estimator defined in Eq.~\ref{eq:acdf_estimator} is upper-bounded by $\|\mathcal{F}\|^2_1$, where $\mathcal{F}\coloneqq |F_0|+\sum_{j=0}^{K} |F_{2j+1}|$. Since the precision $\epsilon = \mathcal{O}(p_0)$ and $F_j=\mathcal{O}(\frac{1}{j})$, the number of samples required is $\mathcal{O}(\|\mathcal{F}\|^2_1p_0^{-2})=\mathcal{O}(\log(K)^2p_0^{-2})$ and the circuit depth is given by $K=\mathcal{O}(\delta^{-1}\log(p_0^{-1}))$.
In Section~\ref{sec:res_spe}, we derive a tight bound for $K$ and show how $\tilde{C}(x)$ can be obtained from expectation values of the form $\bra{\psi_0}T_k(\hat{H})\ket{\psi_0}$ instead of the expectation values of the time propagator of Eq.~\ref{eq:acdf_estimator}.

\section{Methods and Results}
\label{sec:results}

We now describe details of our methodology and summarize both the analytical and numerical results of this work.

\subsection{Molecular Hamiltonians}
\label{sec:molecules}

We investigate four representative molecular systems: Fe$_2$S$_2$, Fe$_4$S$_4$, Co(salophen), and Naphthalene. 
The iron--sulfur clusters Fe$_2$S$_2$ and Fe$_4$S$_4$ are prototypical bioinorganic motifs that exhibit strong electronic correlation and multiple near-degenerate spin states, making them challenging benchmarks for both classical and quantum electronic structure methods. 
Co(salophen) is a transition-metal coordination complex that captures key features of open-shell transition-metal chemistry relevant to catalysis. 
Finally, naphthalene serves as a well-controlled organic $\pi$-conjugated system, providing a complementary moderately correlated reference case that would be a good target for non-trivial experimental quantum comptuing demonstrations.

We employ the active-space Hamiltonians from Ref.~\citenum{li2017spin} for the iron--sulfur clusters and from Ref.~\citenum{Dutkiewicz_2025} for Co(salophen) and naphthalene. 
The corresponding active spaces contain $(30, 20)$, $(54, 36)$, $(26, 27)$, and $(10, 10)$ electrons and spatial molecular orbitals and  for Fe$_2$S$_2$, Fe$_4$S$_4$, Co(salophen), and naphthalene, respectively.

We recall that the eigenenergies $E_k$ of the physical $\hat{\mathcal{H}}$ are related to the eigenvalues $\lambda_k$ of its shifted and normalized counterpart, $\hat{H}$, through $E_k = \beta + \lambda \cdot \lambda_k$.
Different molecules and ways of representating the molecular Hamiltonians lead to different values of the normalization factor $\lambda$.
In this work, we use the state-of-the-art \gls{thc} \gls{bliss}\cite{caesura2025faster} representation of the Hamiltonian.
Its detailed definition is given is Appendix~\ref{sec:norms&shifts}.
The performance of the \gls{qksd} algorithm in other Hamiltonian representations is also studied in Section~\ref{sec:qksd_numerics}. 

We emulate the \gls{qksd} and \gls{spe} algorithms using classically computed \gls{dmrg} or \gls{casci} spectra for the $R$ lowest eigenenergies, which necessarily introduces approximation errors due to the exponential cost associated with exact simulation.
To this end, we approximate the electronic structure of several molecular systems and employ these approximations as proxies for the corresponding exact Hamiltonians.

For instance, the \gls{qksd} matrix elements can be expressed as
\begin{equation}
\begin{split}
\bra{\psi_0} T_k(\hat{H}) \ket{\psi_0}
&= \bra{\psi_0} T_k(\hat{H}) \sum_{r=0}^{N-1} \ket{\lambda_r}\bra{\lambda_r} \ket{\psi_0} \\
&\approx \bra{\psi_0} T_k(\hat{H}) \sum_{r=0}^{R-1} \ket{\lambda_r}\bra{\lambda_r} \ket{\psi_0} \\
&= \sum_{r=0}^{R-1} \left| \braket{\lambda_r | \psi_0} \right|^2 T_k(\lambda_r),\\
&\approx \sum_{r=0}^{R-1} \left| \braket{\tilde{\lambda}_r | \psi_0} \right|^2 T_k(\tilde{\lambda}_r),
\end{split}
\end{equation}
Two approximations are introduced at this stage. 
First, we assume that the initial state $\ket{\psi_0}$ has negligible overlap with the highest $N-R$ eigenstates of the Hamiltonian, so that its support is effectively restricted to the lowest $R$ eigenstates. 
Second, when the exact eigenpairs $\{\ket{\lambda_r}, \lambda_r\}_{r=0}^{R-1}$ are not available, we replace them with eigenpairs $\{\ket{\tilde{\lambda}_r}, \tilde{\lambda}_r\}_{r=0}^{R-1}$ obtained from an approximate procedure.

In particular, for the iron sulfur clusters and Co(salophen) we approximate the first $R=40$ eigenstates using \gls{dmrg} at bond dimension $M=1000$.
We first compute a state-averaged matrix product state, followed by state-specific refinements as implemented in \texttt{Block2}\cite{zhai2023block2}.
At this bond dimension, we obtain ground-state energies of $-116.606$~Ha and $-327.227$~Ha, for Fe$_2$S$_2$ and Fe$_4$S$_4$, respectively.
These energy values are consistent with the literature and it is established that they do not reach chemical accuracy with respect to the exact ground state\cite{Low_2025, sharma2014low}.
Moreover, this bond dimension is insufficient for a quantitatively accurate description of the excited states.
These limitations reflect the constraints of the classical methods in simulating such complex molecules; nevertheless, we assume the resulting spectra qualitatively reflect those of the true systems and are sufficient for assessing the scaling behavior and relative performance of the \gls{qksd} and \gls{spe} algorithms.
To further validate our conclusions, we present additional simulations using the first 500 exact eigenstates of naphthalene obtained with \gls{casci} implementation of \texttt{PySCF}\cite{sun2018pyscf, sun2020recent}.

For the iron sulfur clusters and Co(salophen), 
we construct the spectral populations $\{p_r\}_{r=0}^{R-1}$ by fixing the
ground-state weight $p_0 \in (0,1)$ and distributing the remaining
population exponentially over the excited states.

For $r \ge 1$, we define
\begin{equation}
p_r \coloneqq (1 - p_0)\,
\frac{\e^{-\alpha r}}
{\displaystyle \sum_{k=1}^{R-1} \e^{-\alpha k}},
\qquad \alpha > 0,
\end{equation}
with $p_0$ prescribed exactly. This guarantees normalization, $\sum_{r=0}^{R-1} p_r = 1$. The decaying rate, $\alpha$, controls the concentration of weight in the
low-lying states.

In particular, we construct two sets of initial populations for the iron sulfur clusters: the first with $p_0=0.1$ and $\alpha = 0.001$, and the second with $p_0=0.5$ and $\alpha = 0.1$. 
For Co(salophen), we also choose $p_0 = 0.5$ and $\alpha = 0.1$. 
This procedure yields realistic initial states with initial energy errors of order $\sim 10\,\mathrm{mHa}$.
For naphthalene, the initial state is chosen to be the Hartree-Fock state.
The population vector of the overlaps of the Hartree-Fock state with the first 500 eigenstates is re-normalized to project out the small contribution of the last $63004$ higher lying eigenstates which we do not calculate. 

The \gls{thc}-\gls{bliss} optimization\cite{caesura2025faster} is initialized from the THC tensors obtained with \texttt{OpenFermion}\cite{mcclean2020openfermion} and jointly optimizes the \gls{bliss} parameters ($\alpha_1$, $\alpha_2$ and $\beta_{pq}$) and the \gls{thc} tensors with the Adam optimizer of \texttt{JAX}~\cite{jax2018github}.
The \gls{thc} rank is set to $5n$, where $n$ is the number of molecular orbitals.
The initial values of $\alpha_1$ and $\alpha_2$ are drawn from a Gaussian distribution of width $0.05$, while $\beta_{pq}$ is initialized from a Gaussian of width $0.01$, all centered at zero.
The explicit expressions for the norm and shift are provided in Appendix~\ref{sec:norms&shifts}. 

\subsection{\gls{spe}}
\label{sec:res_spe}

In this section we first show how tighter bounds on the Heaviside approximation can be derived, then how \gls{spe} can be run based off of Chebychev expectation values and finally we benchmark the bound against the accuracy of actually running \gls{spe}.

\subsubsection{Tighter Bounds on The Heaviside Function Approximation Truncation Error}
\label{sec:tight_bounds}
This section presents the derivation of a tighter upper bound between the scaled error function and its truncated expansion used to approximate the Heaviside function $\Theta(x)$.
We recall the Fourier expansion $Q_{\beta, K}(x)$ of the scaled error function $\erf(\sqrt{2\beta}x)\coloneqq \frac{2}{\sqrt{\pi}}\int_0^{\sqrt{2\beta}x}\e^{-t^2}dt$ is given by
\begin{align}
    Q_{\beta, K}(x) = 2\e^{-\beta}\sqrt{2\beta/\pi}\left[I_0(\beta)x + \sum_{j=1}^{K}(-1)^jI_j(\beta)\left(\frac{T_{2j+1}(x)}{2j+1}-\frac{T_{2j-1}(x)}{2j-1}\right)\right], \forall \beta \in \mathbb{R}
\end{align}
where $I_j(\cdot)$ is the modified Bessel function of first kind of order $j$. $\Theta(x)$ is then approximated with $H(x)\coloneqq \frac{Q_{\beta, K}(\sin(x))+1}{2}$ leading to the Fourrier expansion in Eq.~\eqref{eq:Heavside_truncated}.
In Ref.~\citenum{Wan_2022}, Proposition~3 states that
\begin{align} \label{eq:erf_bound}
    \delta = \left|\erf(\sqrt{2\beta}x)-Q_{\beta, K}(x)\right| \leq 2\sqrt{2\beta/\pi}\e^{-\beta}\frac{1}{K} \sum_{j=K+1}^{\infty}I_j(\beta),
\end{align}
and Proposition~4 gives the following upper bound on the right hand side
\begin{align}
    2\e^{-\beta}\sum_{j=K+1}^{\infty}I_j(\beta) \leq 2\e^{-\frac{\left(K+1\right)}{t}^2} + \frac{1}{2}\left(\frac{\e}{t}\right)^t\e^{-\beta}, \quad \forall t\geq \beta.
    \label{eq:rhs_bound}
\end{align}
This allows to bound the value of $K$ given the desired precision $\delta$.

To improve this result, we show how the right hand side of Eq.~\eqref{eq:erf_bound} can be calculated directly instead of bounding it as in Eq.~\ref{eq:rhs_bound}.
Using the generating function of the modified Bessel functions of first kind we have the following identity
\begin{align}
    \sum_{-\infty}^{\infty}I_j(\beta)t^j = \e^{\beta(t+1/t)}, \forall t \in \mathbb{C}.
\end{align}
Therefore, we have
\begin{align}
    \label{eq:tighter_bound}
    \sum_{-\infty}^\infty I_j(\beta) &= \e^{\beta}\\
    \implies 2\sum_{j=K+1}^{\infty} I_j(\beta) &= \e^{\beta} - I_0(\beta) - 2\sum_{j=0}^{K} I_j(\beta) \\
    \implies \sqrt{\frac{2\beta}{\pi}}\frac{2\e^{-\beta}}{K} \sum_{j=K+1}^{\infty}I_j(\beta) & = \sqrt{\frac{2\beta}{\pi}}\frac{1}{K}\left(1-\e^{-\beta}I_0(\beta) - 2\sum_{j=0}^{K} \e^{-\beta}I_j(\beta)\right)
    \label{eq:better_bound}
\end{align}
where we have split the infinite sum in the second line and used the symmetry $I_j(\beta)=I_{-j}(\beta), \forall j \in \mathbb{N}$ and subsequently multiplied by the same factor $\sqrt{\frac{2\beta}{\pi}}\frac{\e^{-\beta}}{K}$ in the third line. 

Figure~\ref{fig:delta_E_vs_K} illustrates the improvements brought by using Eq.~\eqref{eq:better_bound} over Eq.~\eqref{eq:rhs_bound}. It reveals an approximately $2/3$ reduction of $K$ to be promised the same accuracy and that our bound is tight.

\begin{figure}
    \centering
    \includegraphics[width=0.7\linewidth]{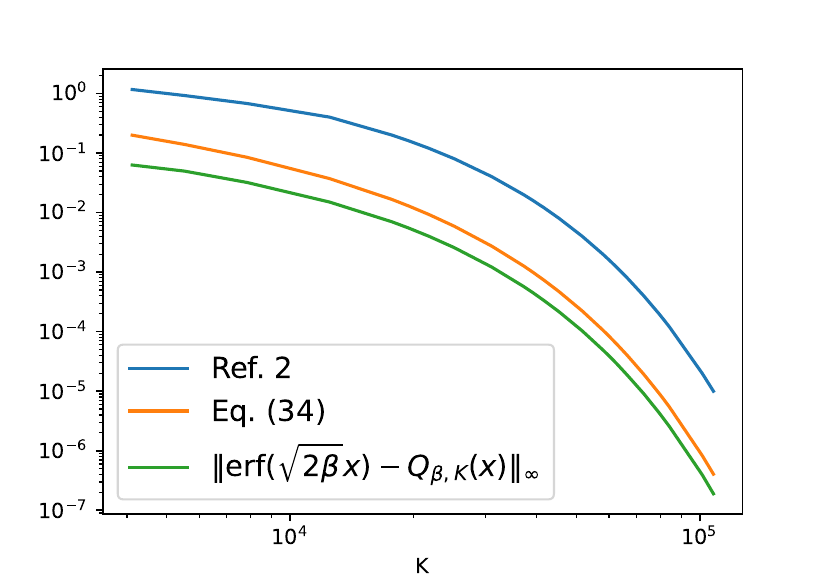}
    \caption{
Bounds on the truncation error of the scaled error function as a function of the truncation order $K$.
Orange is the upper bound from Ref.~\citenum{Wan_2022}.
Blue is the bound from Eq.~\eqref{eq:tighter_bound} and .
green is the actual infinity norm difference between $Q_{\beta,K}$ and the scaled error function $\erf(\sqrt{2 \beta}x)$ estimated numerically by evaluation of both functions on a fine grid.
}
    \label{fig:bounds}
\end{figure}

\subsubsection{Approximate Cumulative Distribution Function from Chebyshev expectation values}
\label{sec:acdf_cheby}
This section demonstrates the derivation of an approximate \gls{cdf} from expectation values with respect to Chebyshev polynomials of the Hamiltonian $\{\langle T_k(\hat{H})\rangle\}_{k=0}^{2K+1}$.
This establishes a direct connection to \gls{qksd}, introduced earlier in Section~\ref{sec:qksd}, and thereby provides a common framework for a meaningful comparison between \gls{spe} and \gls{qksd}. In this setting, both methods rely on the same quantum observables, ${\langle T_k(\hat{H}) \rangle}_{k=0}^{2K+1}$, which are subsequently post-processed classically to estimate the ground-state energy through two distinct procedures.
The comparison then follows naturally, as both methods are characterized by their respective circuit depth, directly reflected in the parameter $K$, and by the total number of circuit repetitions $M$.
 
Moreover, the practical realization of the time-evolution operators, in Eq.~\ref{eq:acdf_estimator}, with methods like Trotter product formulae, produces a total runtime where the scaling is no longer Heisenberg-limited\cite{Lin_2022}.
In Ref.~\citenum{Wan_2022}, a randomized procedure based on sampling from the infinite Taylor expansion of these time-evolution operators restores the optimal scaling; however, it introduces a nonzero—albeit exponentially small—probability of generating exponentially deep circuits. In contrast, our approach, which relies exclusively on Chebyshev polynomials of the Hamiltonian rather than time-evolution operators, avoids these issues entirely.

Let $p_{\pm}(\cdot)$ be the probability distribution over $\{\pm\arccos(\lambda_k)\}_{k=0}^{N-1}$ defined by the squared overlap $p_k\coloneqq |\braket{\psi_0|\lambda_k}|^2$ of the initial state $\ket{\psi_0}$ with the eigenstates of $\hat{H}$
\begin{align}
    p_{\pm}(x)=\sum_{k=0}^{N-1}p_k\delta(x\mp\arccos(\lambda_k))
\end{align}
Let us denote the respective \glspl{cdf} by $C_{\pm}$. Given that \mbox{$\{\arccos(\lambda_k)\geq 0\}_{k=0}^{N-1}$} and $\{-\arccos(\lambda_k)\leq0\}_{k=0}^{N-1}$, this implies that
\begin{align}
    C_{+}(x)+C_{-}(x) &= C_{-}(x) \; \forall x < 0 \label{eq:cheby_acdf}\\
    C_{+}(x)+C_{-}(x) &= 1 + C_{+}(x) \; \forall x > 0 
\end{align}
Now let us denote their approximate \glspl{cdf} with $\tilde{C}_\pm$. With $S_1\coloneqq \{0\}\bigcup\{2j+1\}_{j=0}^{K}$ and using the Heaviside approximation in Eq \ref{eq:Heavside_truncated}, we have 
\begin{align}
    \tilde{C}_{+}(x) &= \left( H \ast p_{+} \right)(x)= \left(\sum_{j\in S_1}F_j\e^{ij(\cdot)} \ast p_{+}(\cdot)\right)(x) \\
    & = \int \sum_{j\in S_1}F_j\e^{ij(x-y)} p_{+}(y)dy\\
    & = \sum_{j\in S_1}F_j\e^{ijx}\int \e^{-ijy} p_{+}(y)dy\\
    & = \sum_{j\in S_1}F_j\e^{ijx} \sum_{k=0}^{N-1}\e^{-ij\arccos(\lambda_k)} p_k\\
    & =\sum_{j\in S_1}F_j\e^{ijx} (\sum_{k=0}^{N-1}\cos(j\arccos(\lambda_k)) p_k-i\sum_{k=0}^{N-1}\sin(j\arccos(\lambda_k)) p_k)\\
    &=\sum_{j\in S_1}F_j\e^{ijx} (\langle T_j(\hat{H})\rangle-i\sum_{k=0}^{N-1}\sin(j\arccos(\lambda_k)) p_k).
\end{align}
Similarly,  
\begin{align}
    \tilde{C}_{-}(x) &=  H \ast p_{-} = \left(\sum_{j\in S_1}\e^{ij(\cdot)} \ast p_{-}(\cdot)\right)(x) \\
    & = \int \sum_{j\in S_1}F_j\e^{ij(x-y)} p_{-}(y)dy\\
    & = \sum_{j\in S_1}F_j\e^{ijx}\int \e^{-ijy} p_{-}(y)dy\\
    & = \sum_{j\in S_1}F_j\e^{ijx} \sum_{k=0}^{N-1}\e^{-ij(-\arccos(\lambda_k))} p_k\\& = \sum_{j\in S_1}F_j\e^{ijx} \sum_{k=0}^{N-1}\e^{ij(\arccos(\lambda_k))} p_k\\
    & =\sum_{j\in S_1}F_j\e^{ijx} (\sum_{k=0}^{N-1}\cos(j\arccos(\lambda_k)) p_k+i\sum_{k=0}^{N-1}\sin(j\arccos(\lambda_k)) p_k)\\
    &=\sum_{j\in S_1}F_j\e^{ijx} (\langle T_j(\hat{H}) \rangle+i\sum_{k=0}^{N-1}\sin(j\arccos(\lambda_k)) p_k)
\end{align}
Summing $\tilde{C}_{+}+\tilde{C}_{-}$, we have
\begin{align}
\label{eq:sum_approx}
    \tilde C_+(x) + \tilde C_-(x) = 2\sum_{j\in S_1}F_j\e^{ijx} \langle T_j(\hat{H}) \rangle
\end{align}

Since the $\arccos(\cdot)$ is a decreasing function, the first discontinuity in $C_{-}(x)$ occurs to the left of the origin at $-\arccos(\lambda_0)$. 
We then propose to use the binary-search presented in Refs.~\citenum{Lin_2022, Wan_2022} to search for the first jump in $\tilde{C}_{-}+\tilde{C}_{+}$ and hence produce an estimate of $\arccos(\lambda_0)$ up to precision $\delta$. Subsequently, $\lambda_0$ is inferred by inverting the $\arccos$ using $\cos$. 

An important remark is in order. The necessary condition in Eq.~\ref{eq:condition_acdf_ok}, which guarantees that the approximate \gls{cdf} remains close to the exact one, holds only if
\begin{equation}
\{-\arccos(\lambda_i)\}_{i=0}^{N-1}
\subset
\left[-\frac{\pi - \delta}{2},\, \frac{\pi - \delta}{2}\right].
\end{equation}
Since $\arccos(\cdot)$ takes values in the interval $[0,\pi]$, this requirement implies
\begin{equation}
\{\arccos(\lambda_i)\}_{i=0}^{N-1}
\subset
\left[0,\, \frac{\pi - \delta}{2}\right],    
\end{equation}
which in turn leads to the necessary condition
\begin{align}
    \label{eq:spe_cond}
    \{\lambda_i\}_{i=0}^{N-1}
    \subset
    \left[\sin\!\left(\frac{\delta}{2}\right),\, 1\right].
\end{align}
This condition can be readily satisfied by expressing $\hat{H}$ as a sum of squares~\cite{Low_2025, rubin2026near}.

Moreover, applying the cosine to retrieve the ground state energy approximation can evidently amplify the accuracy or at worst leave it unchanged depending on the position of the $\lambda_0$ in the interval $\left[\sin(\frac{\delta}{2}), 1\right]$. To illustrate this, let us denote the approximation of $\arccos(\lambda_0)$ with $\tilde \gamma_0$ and, without loss of generality, let us assume that $\tilde{\gamma}_0\geq \arccos(\lambda_0)$. The mean value theorem yield
\begin{align}
    |\cos(\tilde{\gamma}_0) - \cos(\arccos(\lambda_0))| \leq |\sin(\gamma_0)|\delta, \text{ for } \gamma_0 \in \left[\arccos(\lambda_0), \tilde{\gamma}_0 \right].
    \label{eq:error_prop}
\end{align}
Hence the accuracy is in fact amplified when $\lambda_0$ value is near 1.

\subsubsection{Application of the Chebyshev-based \gls{spe} to the ground state of $\mathrm{Fe_4S_4}$}
\label{sec:res_spe_benchmarks}

\begin{figure}
    \centering
    \includegraphics[width=0.7\linewidth]{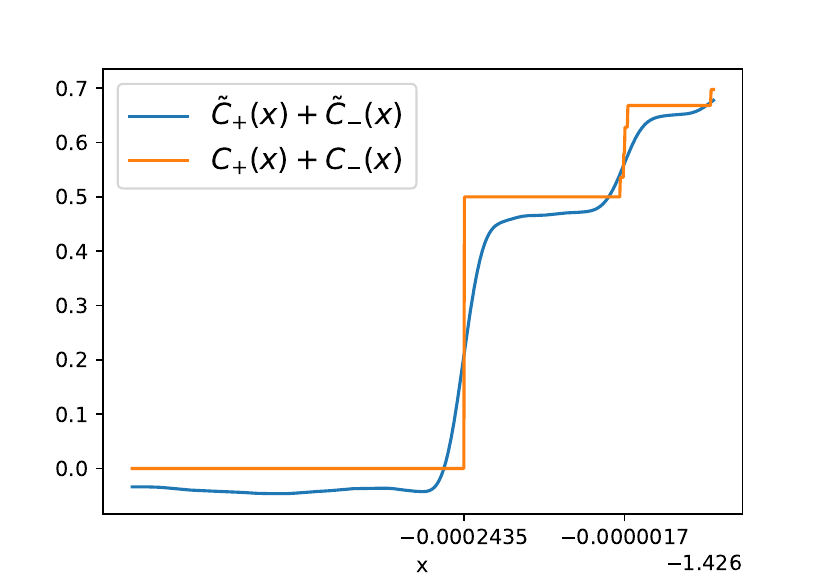}
    \caption{Illustration of $C_{+}+C_{-}$ and its noisy approximation respectively for the DMRG-approximated spectrum of $\mathrm{Fe_4S_4}$ with the jumps, due to lowest five $\arccos(\lambda_k)$ visible.
    The values $\tilde{C}_{+}(x) + \tilde{C}{-}(x)$ for all $x$ were evaluated from a single set of Chebychev expectation values according to Eq.~\eqref{eq:sum_approx} with $K$ and $M$ chosen such that $\delta \leq 5 \times 10^{-5}$. The initial state and $p_0$ are identical to those used in Figure~\ref{fig:fig1} and elsewhere in this work.
    }
    \label{fig:c+c}
\end{figure}

To demonstrate the feasibility of the \gls{spe} implementation proposed in Section~\ref{sec:acdf_cheby}, we show here how to apply it to approximate the ground-state energy of the $\mathrm{Fe_4S_4}$ cluster.
As described in Section~\ref{sec:molecules}, the \gls{spe} algorithm is emulated using the DMRG-approximated spectrum of $\mathrm{Fe_4S_4}$.
We consider the initial-state distribution corresponding to $p_0 = 0.5$ (see Section~\ref{sec:molecules}).
Figure~\ref{fig:c+c} displays the resulting quantity $\tilde{C}_{+} + \tilde{C}{-}$.
It can be observed that $\tilde{C}_{+} + \tilde{C}_{-}$ approximates $C_{-}$ for $x\leq 0$.
$M$ and $K$ are large enough, and thus $\delta$ small enough to make the big jump corresponding to $\arccos(\lambda_0)$ easily identifiable, but smaller jumps due to the two first excited states merge together.
Once can nicely see that values of $\tilde{C}_{+}(x) + \tilde{C}{-}(x)$ are correlated, i.e., the function is smooth, on a scale of $\delta$.

\begin{figure}[htbp]
\centering

\begin{subfigure}{0.45\textwidth}
\centering
\includegraphics[width=\linewidth]{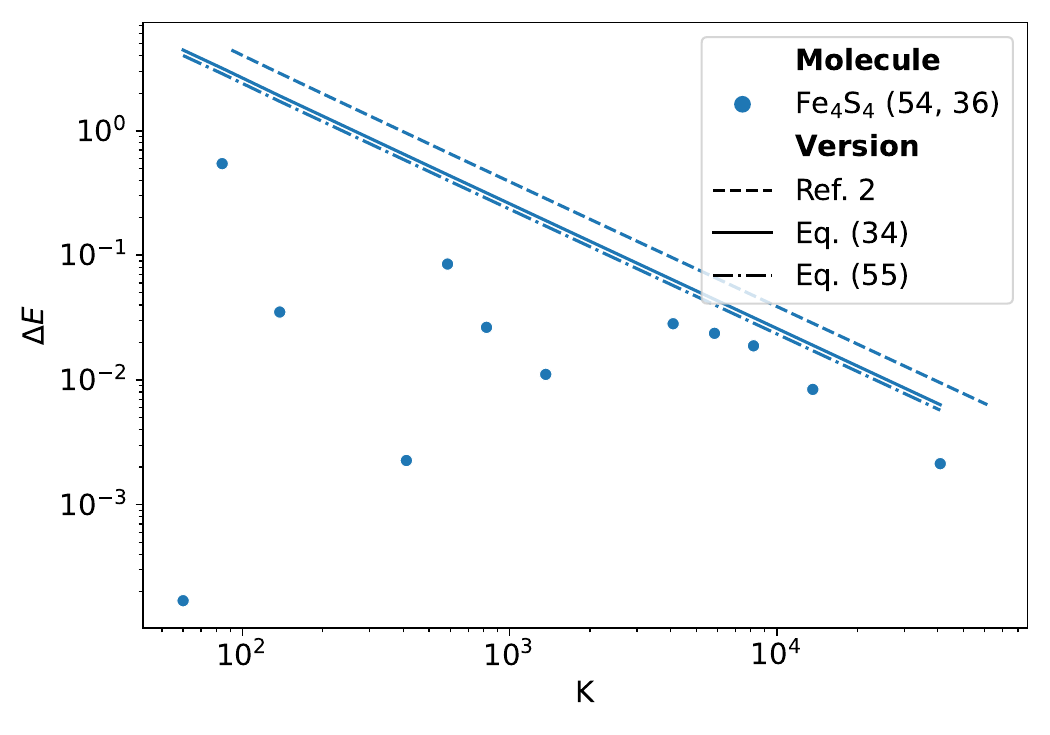}
\caption{}
\end{subfigure}
\hfill
\begin{subfigure}{0.45\textwidth}
\centering
\includegraphics[width=\linewidth]{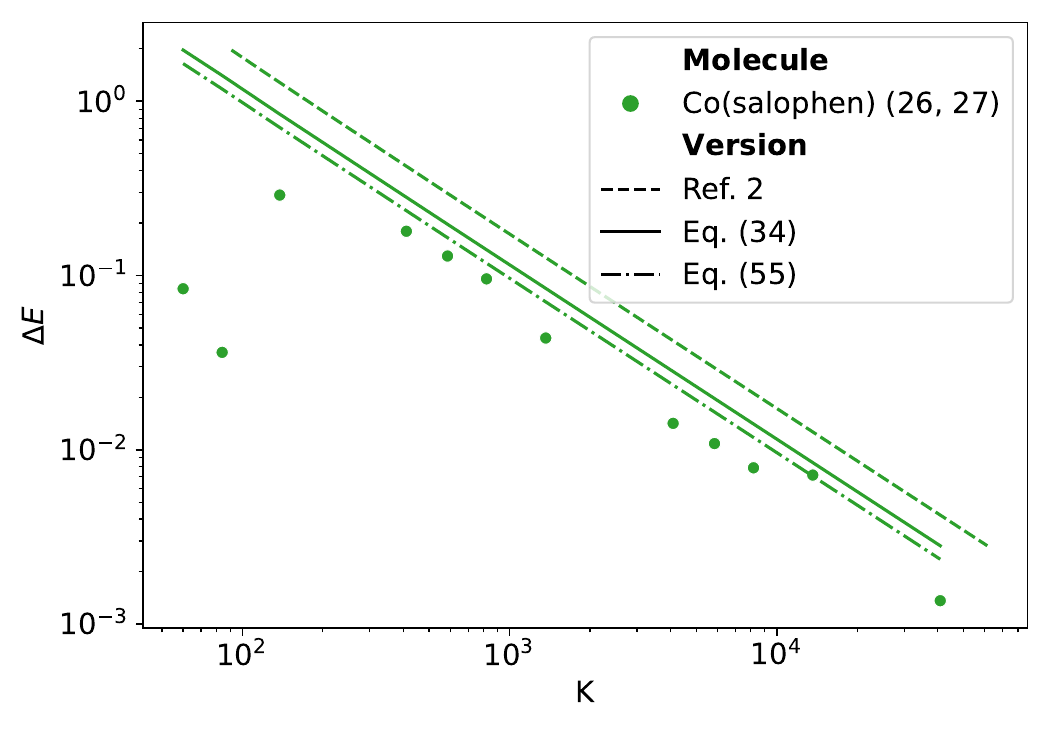}
\caption{}
\end{subfigure}

\vspace{0.5cm}

\begin{subfigure}{0.45\textwidth}
\centering
\includegraphics[width=\linewidth]{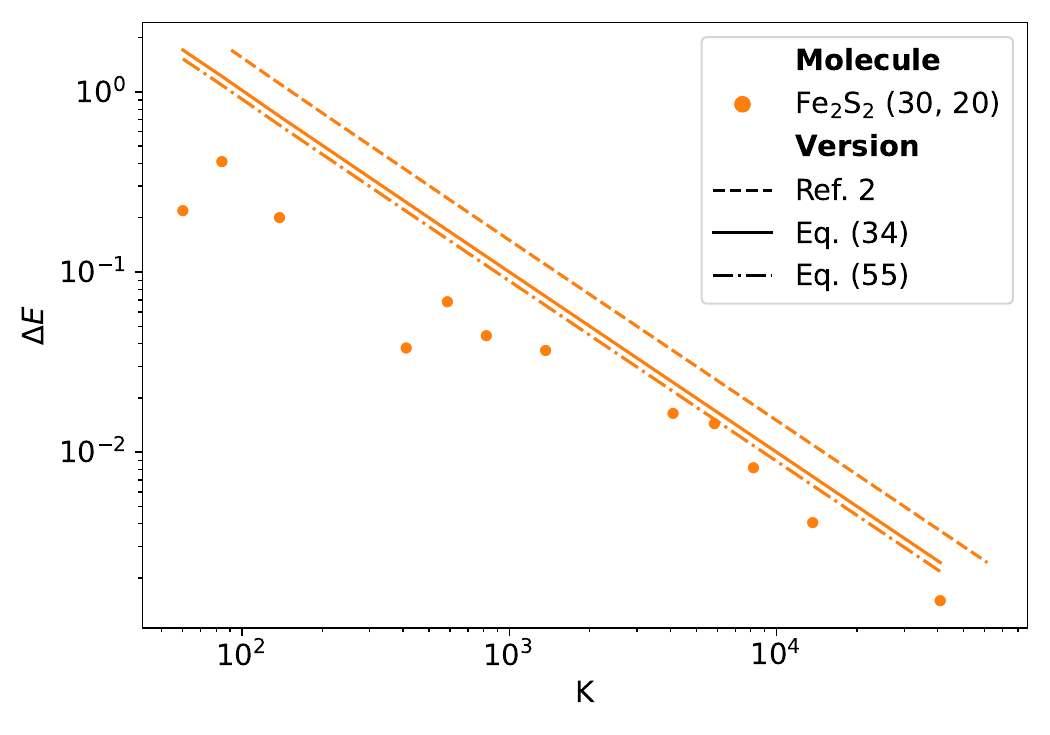}
\caption{}
\end{subfigure}
\hfill
\begin{subfigure}{0.45\textwidth}
\centering
\includegraphics[width=\linewidth]{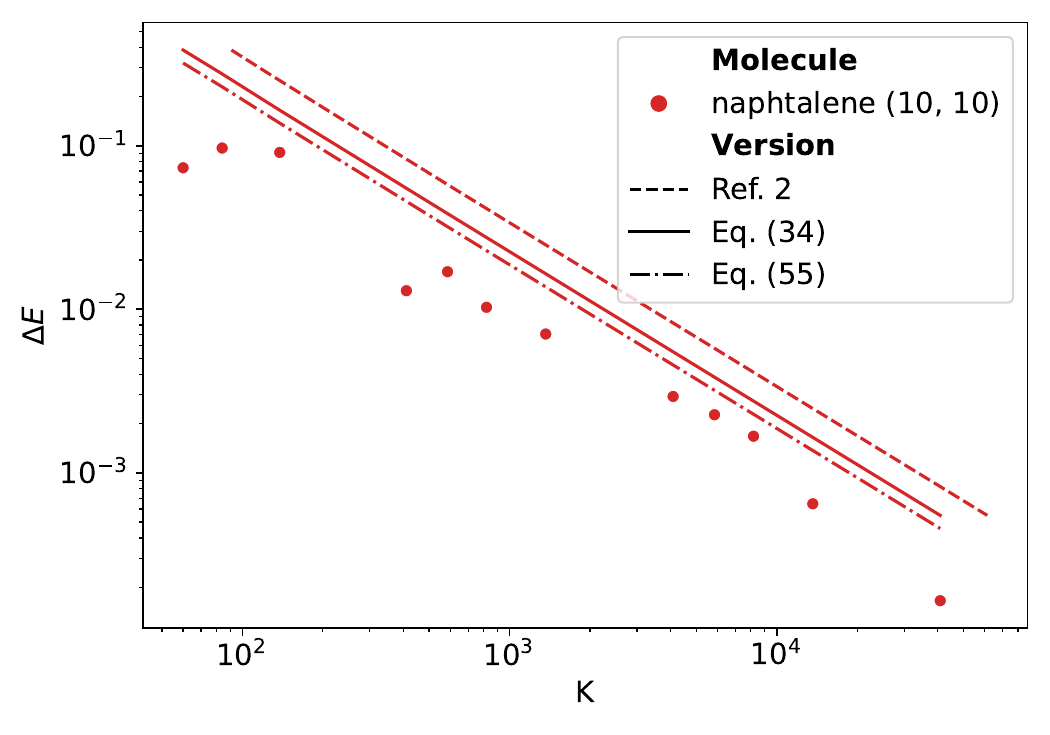}
\caption{}
\end{subfigure}
\caption{\label{fig:delta_E_vs_K}
Comparison of various bounds on the \gls{spe} error $\Delta E$ to the actual \gls{spe} performance as a function of the maximal Chebyshev polynomial degree $K$ for the different molecules considered in this work.
The dashed line is the upper bound from Ref.~\citenum{Wan_2022}.
The solid line is the bound implied by Eq.~\eqref{eq:tighter_bound} and the dash-dotted line is additionally taking into account the error propagation due to spectral amplification according to Eq.~\eqref{eq:error_prop}.
The energy precision actually achieved by \gls{spe} in Hartree is plotted as dots for the $M$ and $K$ inferred from the tightest bound, as was done in Figure~\ref{fig:fig1}
}
\end{figure}

Figure~\ref{fig:delta_E_vs_K} shows the achieved ground-state energy accuracy for our four molecules, namely $\mathrm{Fe_4S_4}$, $\mathrm{Fe_2S_2}$, $\mathrm{Co(salophen)}$ and naphthalene, as a function of the maximal Chebyshev polynomial degree $K$, which is itself set by the target precision $\delta$. The solid line present the upper bound accuracy achieved with $K$ inferred from Eq.~\eqref{eq:tighter_bound} whereas the dashed lines indicates that of Ref. \citenum{Wan_2022} and the dashed-dot shows the effect of the error-propagation in Eq.~\eqref{eq:error_prop}.
The Hamiltonian $\hat{\mathcal{H}}$ is shifted and rescaled by $\beta=\beta_{\text{THC-BLISS}}$ and $2\lambda_{\text{THC BLISS}}$, respectively placing the spectrum in $\left[-\frac{1}{2}, \frac{1}{2}\right]$. Subsequently, $\hat{H}$ is shifted by $-\frac{1}{2}$ and multiplied by (-1) to shift and flip the entire spectrum to $\left[0, 1\right]$. Note that this operation flips the order of the original spectrum and can hence potentially places the ground state energy near the edge of the interval which can result in improved error as demonstrated in Eq.~\eqref{eq:error_prop}.

The total number of samples is chosen as 
\begin{equation}
    M = \frac{2\mathcal{F}^2}{\eta^2},
\end{equation}
where $\eta$ is a lower bound on $p_0$, taken here as $p_0/2$, and $\mathcal{F} = \sum_{j\in S_1} |F_{j}|$. Shot noise resulting from importance sampling is then simulated as follows: for each $\langle T_k(\hat H)\rangle$, a sampling noise is added by drawing from a zero-centered normal distribution with variance $(1-\langle T_k(\hat H)\rangle^2) / M_k$ where $M_k$ is the number of samples allocated, importance sampling according to Eq.~\eqref{eq:sum_approx}, to estimate $\langle T_k(\hat H)\rangle$ and $\sum_k M_k = M$.

\subsection{\gls{qksd}}
This section focuses on the performance of \gls{qksd} for computing the ground-state energy of the iron-sulfur clusters. We first demonstrate that, in the absence of shot noise, it is advantageous to subsample polynomial degrees rather than measuring every degree up to $K$, retaining only those that yield sufficiently non-collinear Krylov vectors. However, we subsequently show that in the presence of shot noise, this subsampling strategy increases the total number of shots required, rendering it impractical.
We also show how the $\lambda$ and $\beta$ parameters affect the performance of the algorithm. 

\subsubsection{Simulations without sampling noise}
\label{sec:qksd_numerics}

\begin{figure}[htbp]
    \centering
    \begin{subfigure}{0.48\textwidth}
        \centering
        \includegraphics[width=\linewidth]{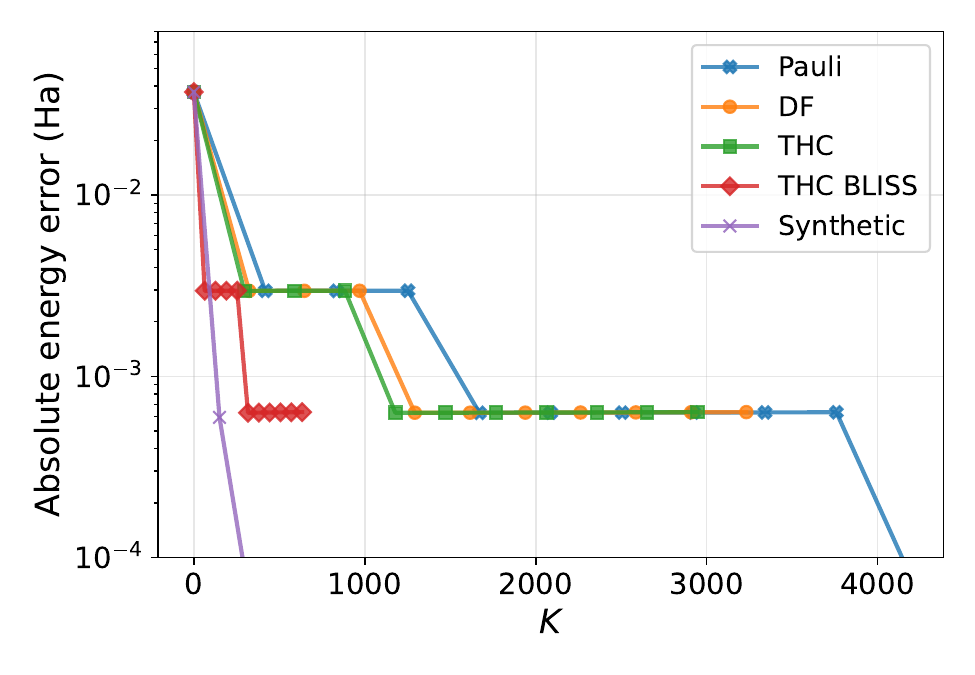}
        \caption{
Fe$_4$S$_4$, $p_0=0.1$
}
    \end{subfigure}
    \hfill
    \begin{subfigure}{0.48\textwidth}
        \centering
        \includegraphics[width=\linewidth]{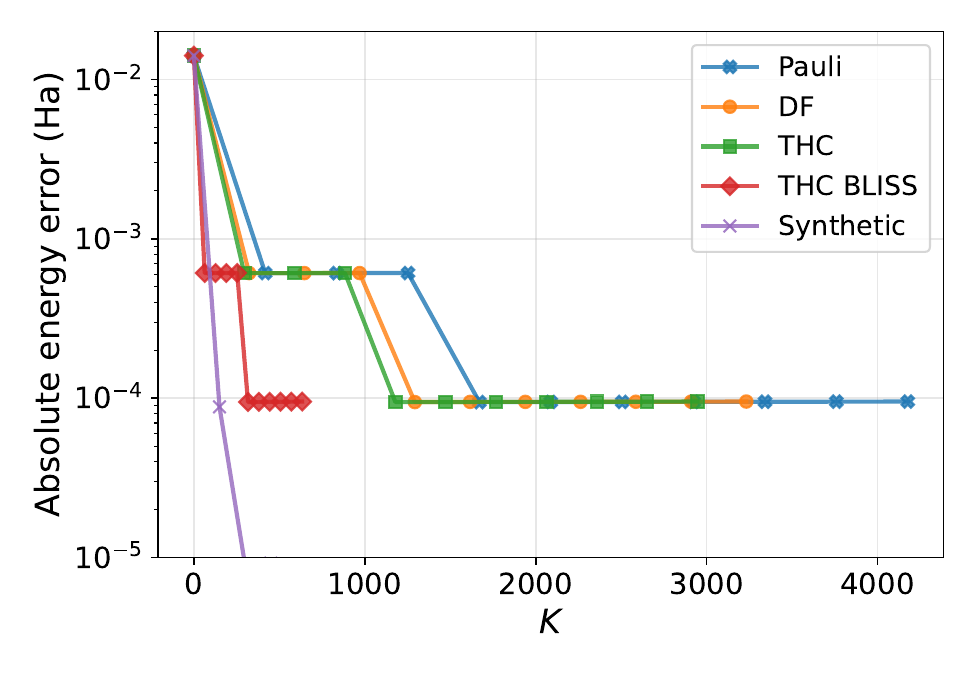}
        \caption{
Fe$_4$S$_4$, $p_0=0.5$
}
    \end{subfigure}
    \medskip
    \begin{subfigure}{0.48\textwidth}
        \centering
        \includegraphics[width=\linewidth]{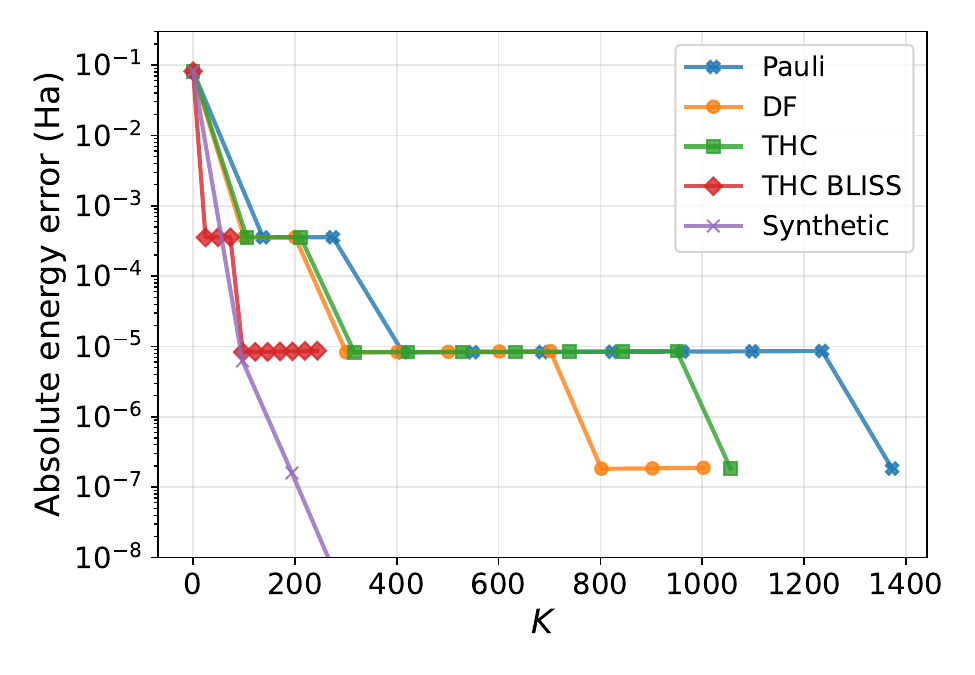}
        \caption{
Fe$_2$S$_2$, $p_0=0.1$
}
    \end{subfigure}
    \caption{
Absolute energy error as a function of $K$, the maximum Chebyshev polynomial order used in the \gls{qksd} algorithm, for different molecules, Hamiltonian representations, and initial-state overlaps.
}
    \label{fig:qksd_allhamiltonians}
\end{figure}

\begin{figure}[htbp]
    \centering
    \begin{subfigure}[t]{0.48\textwidth}
        \centering
        \includegraphics[width=\textwidth]{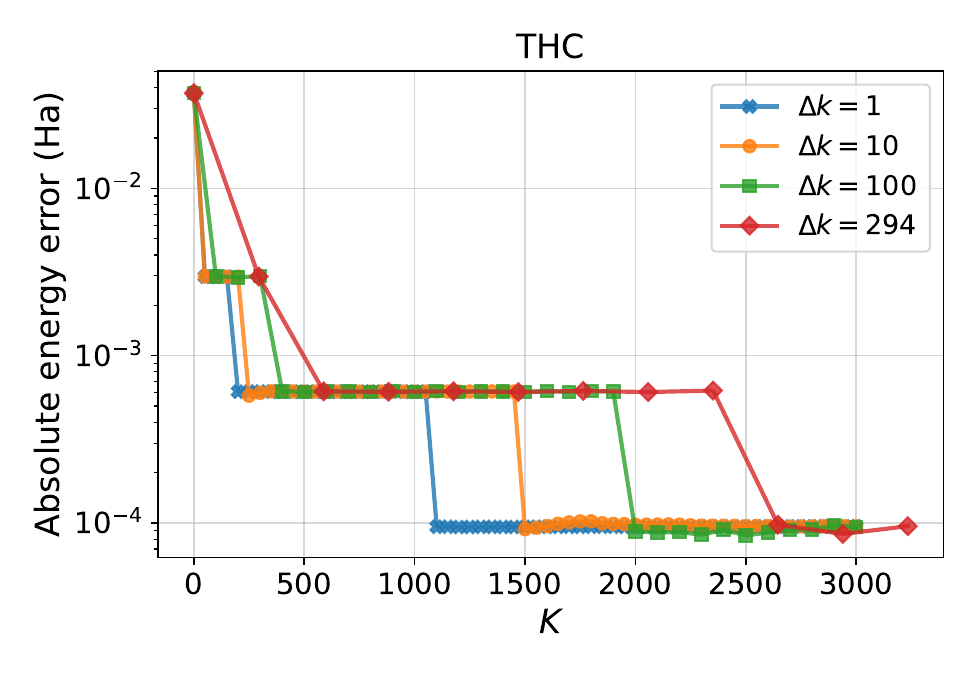}
        \caption{
THC representation
}
    \end{subfigure}
    \hfill
    \begin{subfigure}[t]{0.48\textwidth}
        \centering
        \includegraphics[width=\textwidth]{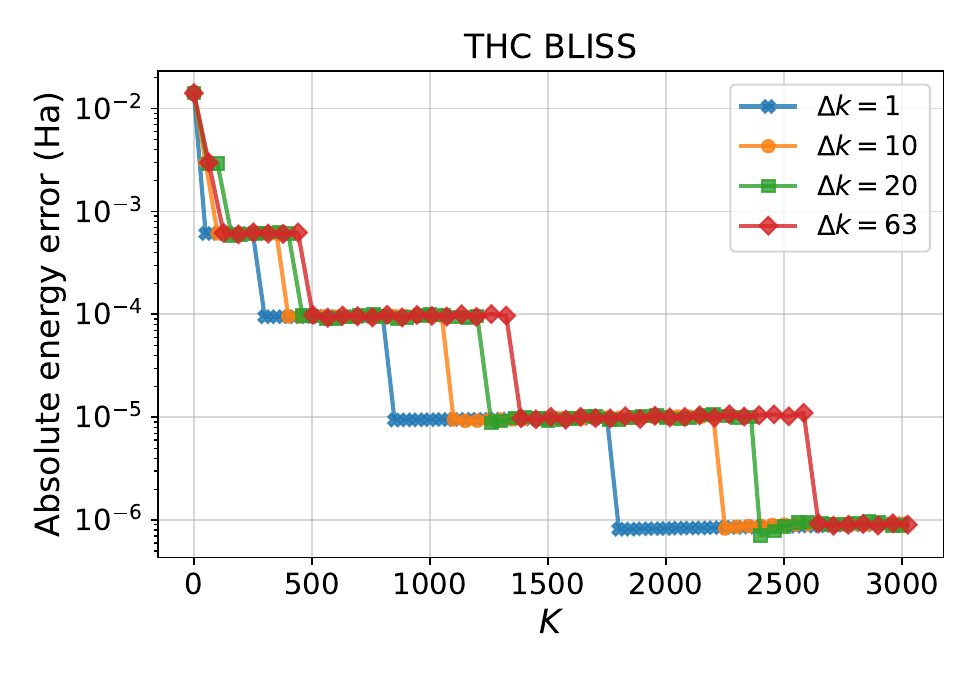}
        \caption{
THC BLISS representation
}
    \end{subfigure}
    \caption{
Absolute energy error as a function of $K$ for Fe$_4$S$_4$ with $p_0 = 0.5$, comparing different polynomial step sizes $\Delta k$.
}
    \label{fig:qksd_all_dks}
\end{figure}

We first analyze the behavior of the \gls{qksd} algorithm in the absence of sampling noise, focusing on the convergence of the ground-state energy as a function of the maximum Chebyshev polynomial order $k$.
The overlap matrix is regularized using a fixed threshold of $10^{-8}$.
Figure~\ref{fig:qksd_allhamiltonians} shows the absolute energy error as a function of the maximal Chebyshev order $K$ for Fe$_4$S$_4$ and Fe$_2$S$_2$ using different Hamiltonian representations and initial-state overlaps.
For each Hamiltonian representation, $K$ is chosen proportional to the corresponding Hamiltonian one-norm $\lambda$, rounded down to the nearest integer, with proportionality factors ranging from $1$ to $10$. 

We compute the Hamiltonian one-norms $\lambda$, the corresponding energy shifts $\beta$, and the renormalized ground-state energies for all Hamiltonian representations considered, and report these quantities in Table~\ref{tab:norms_shifts}.
The Pauli norm and shift are computed following Ref.~\citenum{koridon2021orbital}.
For \gls{df}, we explicitly perform the factorization and evaluate the Burg norm and shift as described in Ref.~\citenum{von2021quantum}.
Tensor hypercontraction is carried out using \texttt{OpenFermion}\cite{mcclean2020openfermion}, with the norm evaluated according to Ref.~\citenum{lee2021even}.
The THC-BLISS optimization\cite{caesura2025faster} is done as explained in Section~\ref{sec:molecules}.
The THC-BLISS norm is evaluated using the same expression as for THC.
All explicit expressions for the norms and shifts are provided in Appendix~\ref{sec:norms&shifts}.
For both \gls{df} and THC, we choose the number of DF leafs and the THC rank to be $5n$, where $n$ is the number of molecular orbitals.

\begin{table}[htbp]
    \centering
    \caption{
Hamiltonian one-norm ($\lambda$), energy shift ($\beta$), and renormalized ground-state energy ($\tilde{E}_0$) for different Hamiltonian representations and molecules.
}
    \label{tab:norms_shifts}
    \begin{tabular}{l|ccc|ccc}
        \hline
        Molecule & \multicolumn{3}{c}{Fe$_2$S$_2$} & \multicolumn{3}{c}{Fe$_4$S$_4$} \\
        \cline{1-4} \cline{5-7}
        Parameter
        & $\lambda$ & $\beta$ & $\tilde{E}_0$
        & $\lambda$ & $\beta$ & $\tilde{E}_0$ \\
        \hline
        Pauli 
        & 137.3 & 101.9 & -0.1069
        & 417.7 & 287.3 & -0.0955 \\
        DF 
        & 100.2 & 52.4 & -0.6409
        & 323.2 & 300.1 & -0.0836 \\
        THC 
        & 105.6 & 105.5 & -0.1048
        & 294.6 & 294.2 & -0.1121 \\
        THC BLISS 
        & 24.3 & 118.9 & 0.0933
        & 63.4 & 336.4 & 0.14404 \\
        Synthetic 
        & 97.0 & 20.0 & -0.9959
        & 150.0 & 178.0 & -0.9948 \\
        \hline
    \end{tabular}
\end{table}

At each $K$, we get the \gls{qksd} $\widetilde{H}$ and $\widetilde{S}$ matrices, defined in Eqs.~\ref{eq:qksd_H} and~\ref{eq:qksd_S} using all Chebyshev polynomial expectation values in $\{\langle T_k(\hat{H})\rangle\}_{k=0}^{2K}$ and solve the generalized eigenvalue problem.
In Figure~\ref{fig:qksd_all_dks}, we observe pronounced plateaus in the energy error as $K$ is increased, across all systems.
These plateaus originate from the quasi-linear dependencies among the Krylov vectors: when several eigenvalues of the overlap matrix fall below the regularization threshold, the effective dimension of the Krylov subspace does not increase and the estimated energy remains relatively unchanged.
Given this behavior it seemed plausible to consider sampling strategy in which polynomial orders are spaced by larger steps $\Delta k > 1$.
In this case, the matrix elements become

\begin{equation}
\begin{split}
\widetilde{H}_{kj} = \tfrac{1}{4} \Big( 
&\langle T_{(k+j)\Delta k+1}(\hat{H}) \rangle 
+ \langle T_{|(k+j)\Delta k-1|}(\hat{H}) \rangle \\
&+ \langle T_{|(k-j)\Delta k + 1|}(\hat{H}) \rangle 
+ \langle T_{|(k-j)\Delta k - 1|}(\hat{H}) \rangle 
\Big)
\end{split}
\end{equation}
and
\begin{equation}
\widetilde{S}_{kj} = \tfrac{1}{2} \Big( \langle T_{(k+j)\Delta k}(\hat{H}) \rangle + \langle T_{|(k-j)|\Delta k}(\hat{H}) \rangle \Big).
\end{equation}
The targeted maximal Chebyshev order remains $K$ but the effective Krylov space dimension becomes $K'=\lfloor K/\Delta k \rfloor$ (hence, the exact maximal Chebyshev order is $K'\Delta k $).
This reduces the number of expectation values to be measured to $2K'$.

The impact of this choice is illustrated in Figure~\ref{fig:qksd_all_dks} for the THC and THC BLISS Hamiltonian representations of Fe$_4$S$_4$ with $p_0 = 0.5$.
While very large values of $\Delta k$ can be suboptimal, since a minimum number of linearly independent Krylov vectors is required to reach a given accuracy, the overall dimension of the Krylov space is nevertheless drastically reduced.
This leads to a significant decrease in the number of polynomials that must be measured on the quantum computer. However, once statistical sampling noise is taken into account, this apparent advantage disappears. As we show in Section~\ref{sec:shot_noise_numerics}, choosing $\Delta k = 1$ ultimately requires a smaller total number of shots, even though it involves measuring a larger number of distinct moments. The reason is that coarser polynomial sampling reduces the magnitude of the retained overlap eigenvalues, thereby increasing the precision required to resolve them above the noise floor. Since the statistical uncertainty scales as $1/\sqrt{M}$, where $M$ is the number of measurements, smaller overlap eigenvalues translate into a quadratically larger shot requirement.

Finally, we note that the effectiveness of larger polynomial steps depends not only on the Hamiltonian one-norm but also on the position of the targeted eigenvalue within the normalized spectrum.
As discussed in Appendix~\ref{sec:qksd_subsampling}, the effective timescale governing the generation of linearly independent Krylov vectors scales as $\lambda \sqrt{1-\lambda_0^2}$, where $\lambda_0$ is the renormalized ground-state energy.
This effect is clearly visible in Figure~\ref{fig:qksd_allhamiltonians}, where a synthetic Hamiltonian, engineered to place $\lambda_0$ close to the spectral edge, exhibits a much steeper improvement of the energy with $K$ and no visible plateaus.

\subsubsection{Simulations with sampling noise}
\label{sec:shot_noise_numerics}

\begin{figure}[htbp]
    \centering
    \begin{subfigure}{0.48\textwidth}
        \centering
        \includegraphics[width=\linewidth]{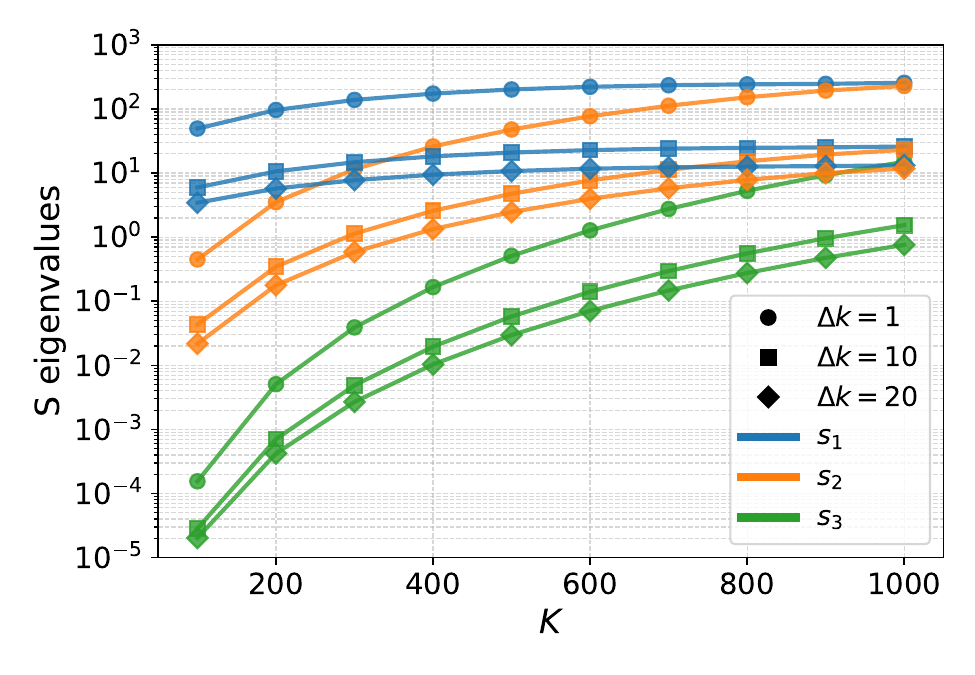}
        \caption{
}
    \end{subfigure}
    \hfill
    \begin{subfigure}{0.48\textwidth}
        \centering
        \includegraphics[width=\linewidth]{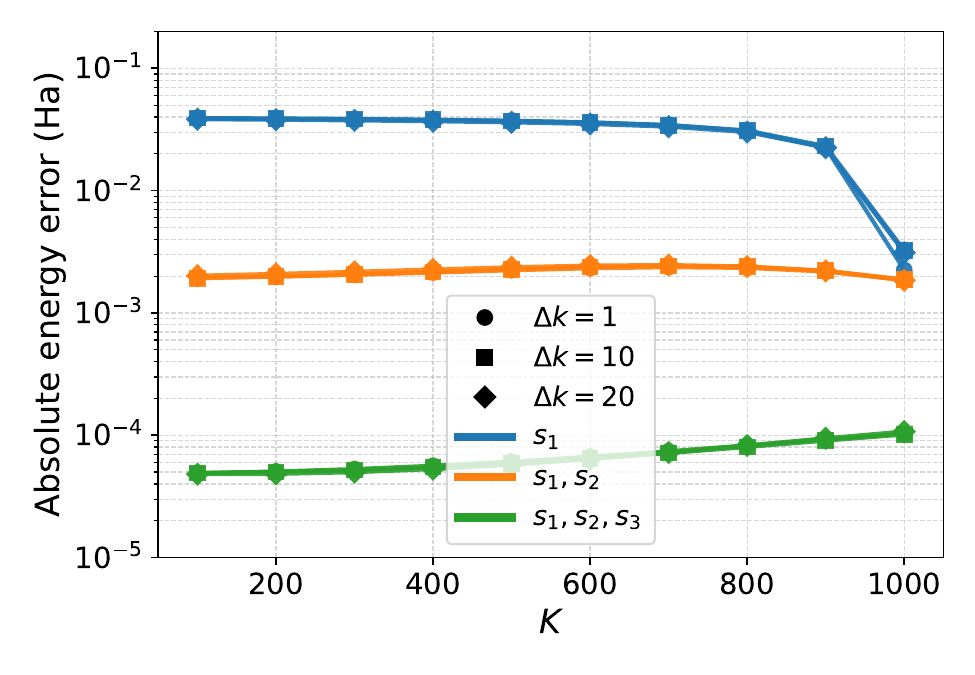}
        \caption{
}
    \end{subfigure}
    \medskip
    \begin{subfigure}{0.66\textwidth}
        \centering
        \includegraphics[width=\linewidth]{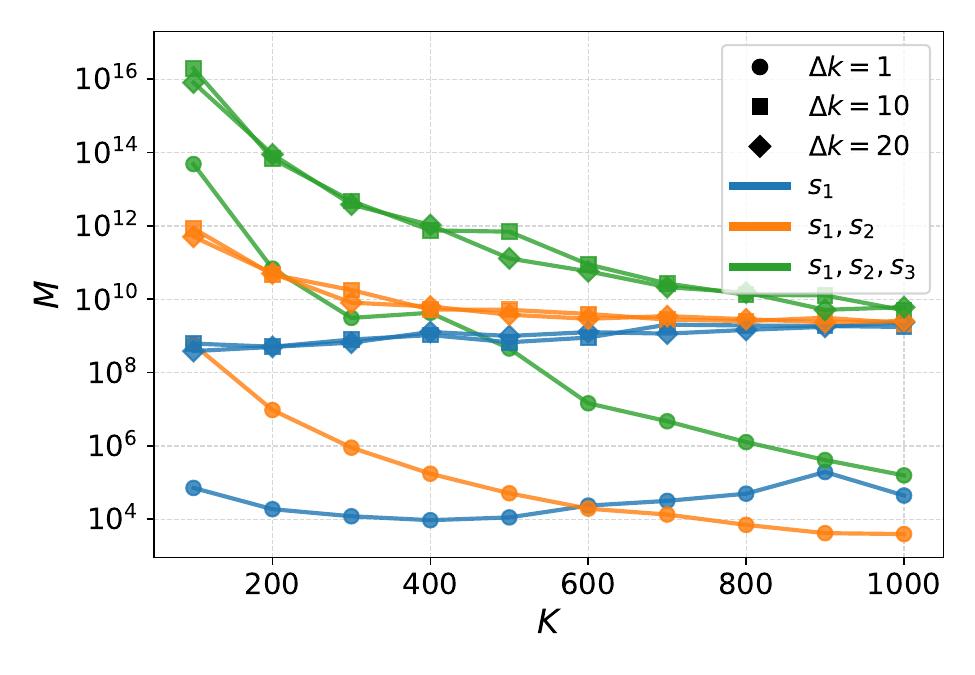}
        \caption{
}
    \end{subfigure}
    \caption{
Fe$_2$S$_2$ with $p_0 = 0.5$: (a) Three largest eigenvalues of the overlap matrix $\tilde{S}$ as a function of the maximal Chebyshev order $K$ for different polynomial step sizes $\Delta k$. (b) Absolute ground-state energy error (with respect to the \gls{dmrg} reference) induced by regularizing the generalized eigenvalue problem and retaining only the largest one, two, or three eigenvalues of $\tilde{S}$. (c) Total number of shots $M$ required to achieve an accuracy of 1 mHa relative to the corresponding noiseless regularized reference. The data illustrate how increasing $\Delta k$ reduces the magnitude of the retained overlap eigenvalues, thereby increasing the sampling cost despite the reduced number of measured moments.
}
    \label{fig:noisy_qksd_fe2s2}
\end{figure}

To model statistical sampling in the \gls{qksd} energy estimator, we allocate a finite shot budget across the Chebyshev expectation values.
Assuming independent measurements, each $\langle T_k(\hat{H}) \rangle \in [-1,1]$ is assigned a Gaussian sampling noise with standard deviation
\begin{equation}
\sigma_k \;=\; \sqrt{\frac{1 - \langle T_k(\hat{H}) \rangle^2}{M_k}},
\end{equation}
where $M_k$ denotes the number of shots used to estimate $\langle T_k(\hat{H}) \rangle$ and $M = \sum M_k$.

We base the shot allocation on the sensitivity of the estimated ground-state energy $E_0$ to each Chebyshev moment.
Concretely, we compute the gradient vector $\mathbf{g}$ with components $g_k = \partial E_0/\partial \langle T_k(\hat{H}) \rangle$ by automatic differentiation (using \texttt{JAX}\cite{jax2018github}) of the classical post-processing routine that maps $\{\langle T_k(\hat{H}) \rangle\}_{k=0}^{2k+3}$ to the regularized generalized eigenvalue estimate.
Given a total shot budget $M$, we distribute shots proportionally to the absolute gradient magnitude,
\begin{equation}
M_k \;=\; \left\lfloor \frac{|g_k|}{\sum_j |g_j|}\, M \right\rfloor,
\end{equation}
and enforce a minimum of one shot for all measured moments, $M_k \leftarrow \max(M_k,1)$.

In simulation we do this by evaluating $g_k$ at the noiseless $\langle T_k(\hat{H}) \rangle$ whereas in a real hardware experiment a part of the shot budget can be used to estimate the $\langle T_k(\hat{H}) \rangle$ and subsequently distribute more shots based on $g_k$.
This should allow one to get very close to the distribution implied by the exact expectation values.
In our implementation, the moments $\langle T_0(\hat{H}) \rangle$ and $\langle T_1(\hat{H}) \rangle$ are excluded from sampling and we set $M_0=M_1=0$, since they are known exactly.
The resulting $\{M_k\}$ determine the noise levels $\{\sigma_k\}$ used to generate noisy moments, which are then propagated through the \gls{qksd} post-processing to estimate the induced error on $E_0$.

The regularization of the overlap matrix $\widetilde{S}$ plays a central role in determining the sampling requirements of the \gls{qksd} algorithm.
In practice, all eigenvalues of $\widetilde{S}$ that cannot be reliably distinguished from statistical noise must be discarded, as retaining such ill-resolved directions leads to severe instabilities in the generalized eigenvalue problem.
Consequently, the total number of shots required is governed by the magnitude of the smallest eigenvalue of $\widetilde{S}$ that is retained after regularization: the noise level must be sufficiently low to resolve this eigenvalue with adequate precision.

To make this dependence explicit, we analyze the noiseless overlap spectra for Fe$_2$S$_2$ with $p_0=0.5$, focusing on the three largest eigenvalues of the overlap matrix $\widetilde{S}$ as functions of the maximal Chebyshev order $K$ and the polynomial step size $\Delta k$ in Figure~\ref{fig:noisy_qksd_fe2s2}(a).
The $\widetilde{S}$ matrix eigenvalues increase monotonically with $K$, reflecting the growing linear independence of the Krylov vectors as the subspace is expanded (see Appendix~\ref{sec:appendix_seigs_growth} for a rigorous analysis of the evolution of the eigenvalues with $K$).
For a fixed value of $K$, we observe a systematic ordering of the overlap eigenvalues with respect to the step size: the eigenvalues obtained for $\Delta k=1$ are always larger than those obtained for $\Delta k=10$, which in turn exceed those obtained for $\Delta k=20$.
This ordering is a direct consequence of the Poincaré separation theorem, which states that the eigenvalues of a principal subspace interlace those of the full space, implying that coarser sampling in polynomial order necessarily leads to smaller overlap eigenvalues.

Using the same noiseless data, we then quantify the error introduced by regularization alone by solving the generalized eigenvalue problem while retaining only the largest one, two, or three eigenvalues of $\widetilde{S}$.
The resulting absolute energy errors are reported relative to the \gls{dmrg} reference ground-state energy in Figure~\ref{fig:noisy_qksd_fe2s2}(b).
This analysis isolates the intrinsic bias induced by truncating the Krylov subspace, independently of any statistical noise, and highlights the trade-off between numerical stability and achievable accuracy.

Finally, we connect the magnitude of the retained overlap eigenvalues to the shot complexity required for a target accuracy.
For each regularization threshold, we estimate the total number of shots needed to reach $1\,\mathrm{mHa}$ accuracy with respect to the noiseless reference (obtained with the same regularization threshold).
Shot noise is introduced stochastically over $100$ independent trials.
The required shot budget is determined iteratively, starting from an initial estimate
\begin{equation}
M_{\mathrm{guess}} = \frac{1}{(s\,\varepsilon)^2},
\end{equation}
where $s$ is the smallest retained eigenvalue of $\widetilde{S}$ and $\varepsilon$ is the target accuracy, and then refined using multiplicative decimal steps until the desired success criterion is met (average error below 1 mHa with respect to noiseless reference over 100 independent trials). 

The resulting number of shots is shown in Figure~\ref{fig:noisy_qksd_fe2s2}(c).
Although increasing $\Delta k$ reduces the number of measured Chebyshev moments, it simultaneously decreases the magnitude of the retained eigenvalues of the overlap matrix $\widetilde{S}$, as shown in Figure~\ref{fig:noisy_qksd_fe2s2}(a).
Since the generalized eigenvalue problem must be regularized by discarding eigenvalues below the sampling noise floor, smaller retained $\widetilde{S}$ eigenvalues impose a significantly stricter precision requirement.
Because statistical fluctuations scale as $1/\sqrt{M}$, resolving these eigenvalues requires a quadratically larger total number of shots $M$.
Consequently, the apparent reduction in the number of measured moments for $\Delta k>1$ does not translate into a lower sampling cost.

\begin{figure}
    \centering
    \includegraphics[width=0.7\linewidth]{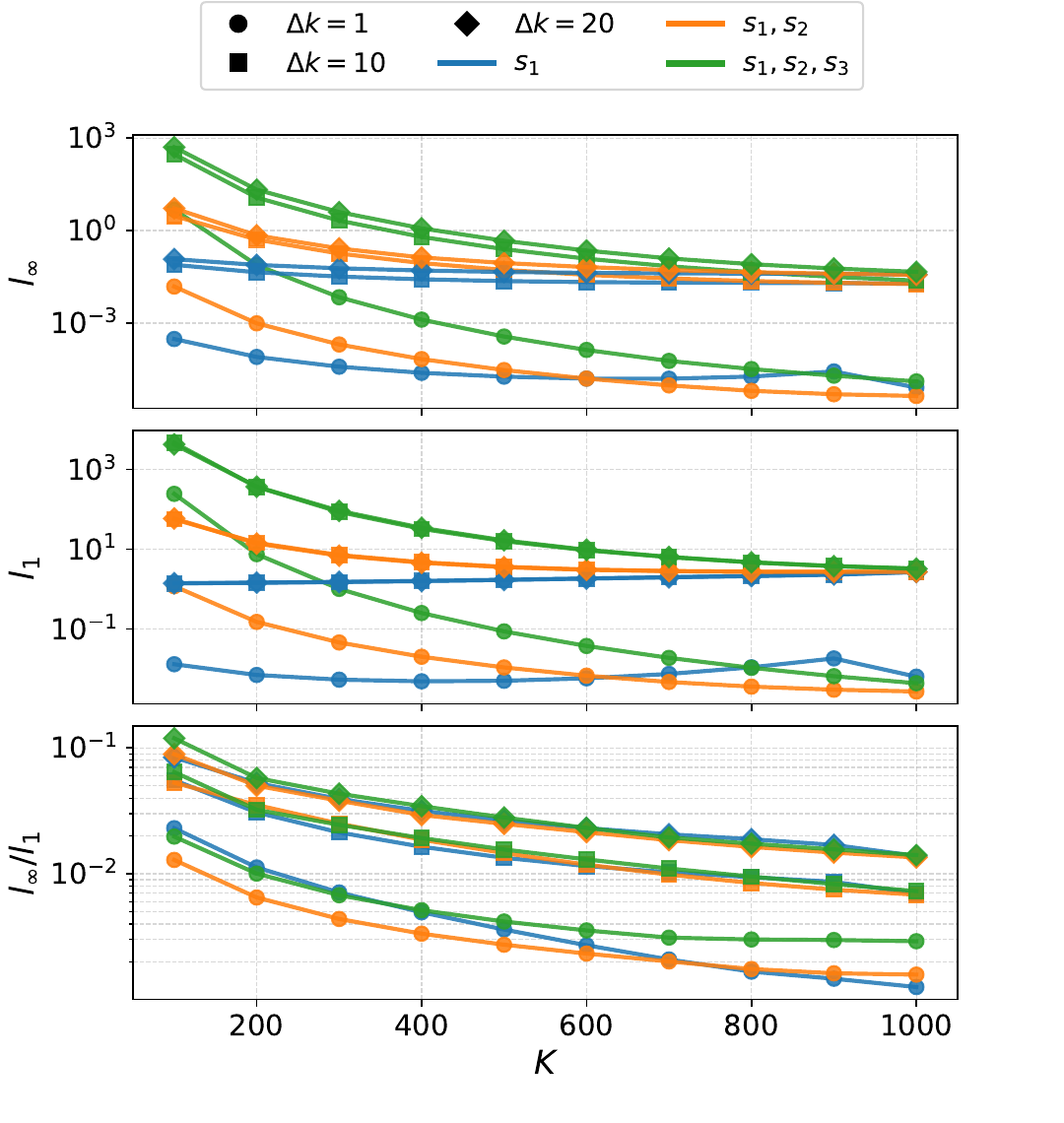}
    \caption{
Fe$_2$S$_2$, $p_0=0.5$: $\ell_\infty$ and $\ell_1$ norms of the energy gradient $g_k \coloneqq \partial E_0 / \partial T_k$ as a function of the Krylov dimension $K$ for different values of $\Delta k$ and for increasing regularization levels i.e. retaining the largest ($s_1$), the two largest ($s_1,s_2$) and the three largest ($s_1,s_2,s_3$) overlap matrix eigenvalues.
}
    \label{fig:g_norms}
\end{figure}

\begin{figure}
    \centering
    \includegraphics[width=\linewidth]{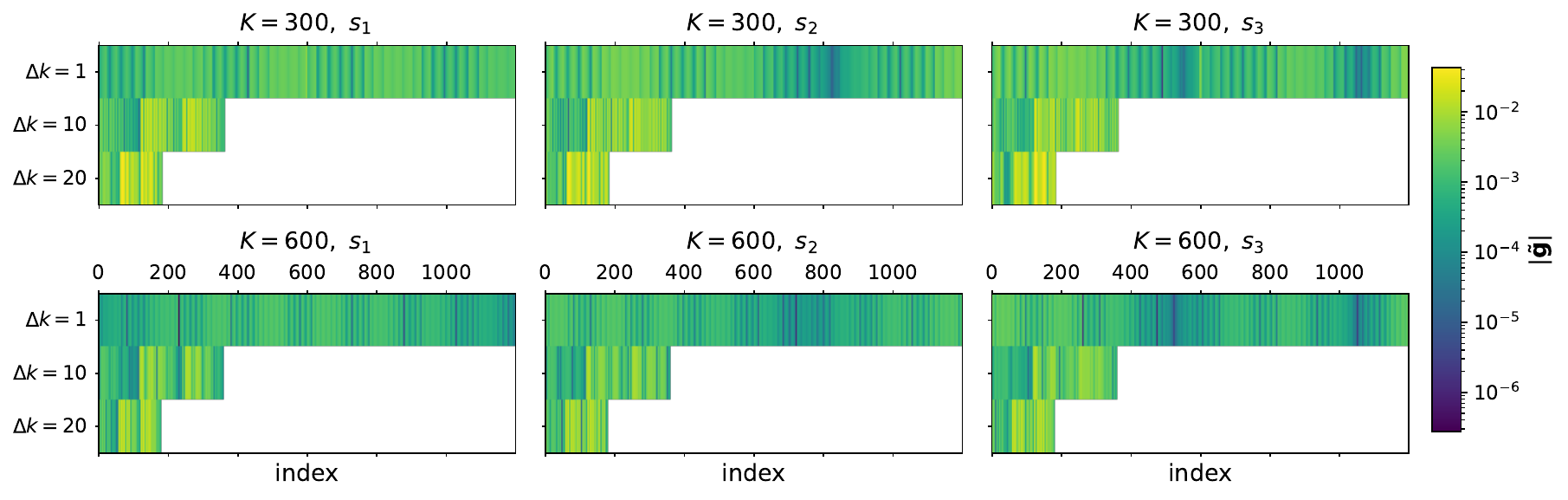}
    \caption{
Normalized gradient $\tilde{\mathbf{g}}=\mathbf{g}/\|\mathbf{g}\|$ used for shot allocation for Fe$_2$S$_2$ with $p_0=0.5$ as a function of the index of the eigenvalues of the overlap matrix $\widetilde S$ at different $K$ values.
}
    \label{fig:fe2s2_gradient}
\end{figure}

The effect of increasing $\Delta k$ on the sampling cost can further be understood by analyzing the gradient $g_k = \partial E_0 / \partial \langle T_k(\hat{H})\rangle$.
Figure~\ref{fig:fe2s2_gradient} shows the normalized gradients used for shot allocation at two different values of $K$.
Because each gradient is normalized by its $\ell_1$ norm, these heatmaps visualize the relative distribution of sensitivity among Chebyshev moments.
They show that for $\Delta k>1$ the sensitivity becomes increasingly concentrated on a smaller subset of low-index moments, whereas for $\Delta k=1$ the sensitivity is distributed more uniformly across the available moments.
This indicates that coarse polynomial sampling leads to a greater reliance on fewer, more critical observables.

The absolute impact of statistical noise is captured by the gradient norms shown in Figure~\ref{fig:g_norms}, which reports the $\ell_\infty$ and $\ell_1$ norms of the gradient as a function of $k$ for different $\Delta k$ and regularization levels.
In contrast to the normalized heatmaps, Figure~\ref{fig:g_norms} shows that both norms are systematically larger for $\Delta k>1$, demonstrating that coarse polynomial sampling amplifies the absolute sensitivity of the estimated energy to statistical fluctuations in the measured moments.
In addition, the ratio $\ell_\infty / \ell_1$ is significantly larger for $\Delta k>1$ than for $\Delta k=1$, demonstrating that coarse polynomial sampling not only increases the global sensitivity of the estimator but also enhances the relative dominance of the most sensitive moments.
Additional numerical results for Fe$_4$S$_4$, Co(salophen) and naphthalene, demonstrating the same qualitative behavior of the overlap spectra, gradient structure, and sampling cost, are provided in Appendix~\ref{sec:appendix_qksd_noise}.

\section{Conclusions}
\label{sec:conclusion}
We have optimized and compared two early fault-tolerant methods for the computation of molecular eigenenergies, namely \acrfull{qksd} and \acrfull{spe}.
To that end, we have developed an \gls{spe} variant that invokes qubitized block encoding walk operators instead of time evolution operators.
This variant uses expectation values with respect to Chebyshev polynomials $\langle T_k(\hat{H})\rangle$ of the Hamiltonian $\hat{H}$ up to a maximum degree $K$ instead of time evolution as input.

In this formulation, the total runtime of \gls{spe} actually achieves Heisenberg scaling and the maximal polynomial degree $K$ --and hence the circuit depth-- scales linearly with the inverse target precision $\delta$, i.e., $K \in \mathcal{O}(\delta^{-1})$.
Furthermore, in this setting, \gls{spe} estimates the arc-cosine of the ground state energy and can therefore potentially benefit from the error propagation when inverting the arc-cosine, as is the case for \acrfull{qpe} \cite{Poulin_2018, babbushEncoding2018} with qubitized walk operators.

We have further improved the upper bound on the truncation error of the scaled error function, achieving an of one order of magnitude enhancement compared to the original result \cite{Wan_2022} and  have thereby reduced of the truncation order $K$ by a factor $2/3$ to guarantee relevant precisions $\delta$.

The choice of the Chebyshev polynomial basis to generate the Krylov subspace enables the computation of the quadratically many entries of the overlap and projected Hamiltonian matrix from linearly many expectation values and reduces the maximum circuit depth, to compute said expectation values, by a factor of $2$ (compared to Hadamard tests). 
For Hamiltonians with large norm, the Krylov states $T_k(\hat{H}) |\psi_0\rangle$ for close-by $k$ become quasi-linear.
We have examined subsampling $k$ spaced by $\Delta k$ as a remedy and have derived the potentially optimal subsampling rate $\Delta k$.
This is effective in the noiseless case; however, in the presence of shot-noise, we discovered that using subsampling requires a larger total number of shots $M$ compared to a more optimal distribution of shots over all $k$ and can explain this behavior from properties of the overlap matrix.
We propose to use automatic differentiation through the \gls{qksd} diagonalization routine to determine the sensitivity of the final output energy on the individual Chebychev expectation values and to distribute shots accordingly.


We benchmarked and compared the total number of shots $M$ as well as the respective circuit depth in terms of the maximal Chebyshev polynomial degree $K$ required by both methods to reach comparable, relevant accuracies.
We find that \gls{spe} generally needs larger $K$ than \gls{qksd}.
For shallow circuits, characterized by small $K$, \gls{qksd} however requires a prohibitively large total number of samples $M$.
The required $M$ however exhibits a pronounced decay with increasing $K$, yielding a number of samples comparable to that of \gls{spe} with still respectively shallower circuits. 
While no circuit level noise was considered in this work, one thing such noise does is put an upper bound on the maximal feasible $K$.
This makes \gls{qksd} look more attractive on hardware with only weak or no error correction because it can be run with $K\leq 10^{4}$, however, even with that method, there is a steep tradeoff between the number of repetitions $M$ and the maximal feasible depth $K$.
Overall, we find that close to $10^5$ shots seem to be necessary to successfully simulate any of the considered Hamiltonians.

A way to improve over this, which will be explored in follow-up work, is to combine \gls{qksd} and \gls{spe} with the latest generation of Hamiltonian factorization techniques, namely DFTHC\cite{rubin2026near, Low_2025}.
This has the potential to further lower the Hamiltonian norms, the non-Clifford count to reach a certain $K$, and unlock the potential of both methods to truly profit from spectral amplification thereby lowering $K$.

We acknowledge discussions with Maximilian Scheurer, J\'{e}r\^ome F.\ Gonthier, and Ammar Kirmani.
QC Ware Corp.\ acknowledges generous funding from Covestro for the undertaking of this project and P.J.O. owns stock/options in QC Ware Corp.
Covestro acknowledges funding from the German Bundesministerium für Forschung, Technologie und Raumfahrt (BMFTR) through project PASQUOPS (13N17250).

\clearpage
\appendix
\begin{appendices}

\section{Norms and shifts in different Hamiltonian representations}
\label{sec:norms&shifts}

The molecular electronic Hamiltonian in second quantization reads
\begin{equation}
\hat{\mathcal{H}} =
\sum_{\sigma \in \{\uparrow,\downarrow\}}
\sum_{p,q=0}^{n-1}
h_{pq}\,
\hat{a}_{p,\sigma}^{\dagger}
\hat{a}_{q,\sigma}
+
\frac{1}{2}
\sum_{\alpha,\beta \in \{\uparrow,\downarrow\}}
\sum_{p,q,r,s=0}^{n-1}
g_{pqrs}\,
\hat{a}_{p,\alpha}^{\dagger}
\hat{a}_{r,\beta}^{\dagger}
\hat{a}_{s,\beta}
\hat{a}_{q,\alpha}.
\label{eq:ham_convention1}
\end{equation}
Here $\hat{a}^{\dagger}_{p,\sigma}$ and $\hat{a}_{p,\sigma}$ are fermionic creation and annihilation operators associated with the $p^{\mathrm{th}}$ spatial orbital and spin $\sigma$.
The number of spatial orbitals is $n$.
The one- and two-electron integrals are defined as
\begin{equation}
h_{pq}
=
(p | h | q)
=
\int d\mathbf{r}\;
\phi_p(\mathbf{r})
\left(
-\frac{1}{2}\nabla^2
- \sum_I \frac{Z_I}{r_I}
\right)
\phi_q(\mathbf{r}),
\end{equation}
and
\begin{equation}
g_{pqrs}
=
(pq | rs)
=
\iint d\mathbf{r}_1 \, d\mathbf{r}_2 \,
\phi_p(\mathbf{r}_1)\phi_q(\mathbf{r}_1)
\frac{1}{r_{12}}
\phi_r(\mathbf{r}_2)\phi_s(\mathbf{r}_2),
\end{equation}
where we assume real spatial orbitals $\phi_p$.

For the factorization techniques considered below, it is convenient to rewrite the Hamiltonian in a form where the two-body operator is expressed as a product of one-body operators.
After reordering fermionic operators, the Hamiltonian becomes
\begin{equation}
\begin{split}
\hat{\mathcal{H}}
&=
\sum_{\sigma}
\sum_{p,q}
t_{pq}\,
\hat{a}_{p,\sigma}^{\dagger}
\hat{a}_{q,\sigma}
+
\frac{1}{2}
\sum_{\alpha,\beta}
\sum_{p,q,r,s}
g_{pqrs}\,
\hat{a}_{p,\alpha}^{\dagger}
\hat{a}_{q,\alpha}
\hat{a}_{r,\beta}^{\dagger}
\hat{a}_{s,\beta} \\
&=
\sum_{p,q}
t_{pq}\,
\hat{E}_{pq}
+
\frac{1}{2}
\sum_{p,q,r,s}
g_{pqrs}\,
\hat{E}_{pq}
\hat{E}_{rs},
\end{split}
\label{eq:ham_convention2}
\end{equation}
where
\begin{equation}
t_{pq}
=
h_{pq}
-
\frac{1}{2}
\sum_{r=0}^{n-1}
g_{prrq},
\end{equation}
and
\begin{equation}
\hat{E}_{pq}
=
\sum_{\sigma}
\hat{a}_{p,\sigma}^{\dagger}
\hat{a}_{q,\sigma}.
\end{equation}

\subsection{Pauli representation}

In this representation, the Hamiltonian is taken as written in Eq.~\eqref{eq:ham_convention1} and mapped to Pauli operators using the Jordan--Wigner transformation.
The one-norm and shift are then given by\cite{koridon2021orbital}
\begin{align}
\beta_{\mathrm{Pauli}}
&=
\left|
E_{\mathrm{nuc}}
+
\sum_{p}
h_{pp}
+
\frac{1}{2}
\sum_{p r}
g_{p p r r}
-
\frac{1}{4}
\sum_{p r}
g_{p r r p}
\right|,
\\[1em]
\lambda_{\mathrm{Pauli}}
&=
\sum_{p q}
\left|
h_{p q}
+
\sum_{r}
g_{p q r r}
-
\frac{1}{2}
\sum_{r}
g_{p r r q}
\right|
+
\frac{1}{2}
\sum_{\substack{p > r \\ s > q}}
\left|
g_{p q r s}
-
g_{p s r q}
\right|
+
\frac{1}{4}
\sum_{p q r s}
\left|
g_{p q r s}
\right|.
\end{align}

\subsection{Double factorization}

In the double-factorization (DF) representation\cite{berry2019qubitization, motta2021low, kivlichan2018quantum}, the two-electron integrals are approximated as
\begin{equation}
g_{pqrs}
\approx
\sum_{t=1}^{N_{\mathrm{DF}}}
\sum_{k,l=0}^{n-1}
U^{t}_{pk}
U^{t}_{qk}
V^{t}_{kl}
U^{t}_{rl}
U^{t}_{sl},
\end{equation}
where each matrix $\bm{U}^t$ is orthogonal.
$N_{\mathrm{DF}} \leq n^2$ is the number of leafs and is chosen according to the desired accuracy with $\mathcal{O}(n)$.

Using this decomposition, the Hamiltonian can be rewritten as
\begin{equation}
\hat{\mathcal{H}}
=
\sum_{p,q}
t_{pq}\,\hat{E}_{pq}
+
\frac{1}{2}
\sum_{t=1}^{N_{\mathrm{DF}}}
\sum_{k,l}
V^{t}_{kl}\,
\hat{G}_t^{\dagger}
\hat{E}_{kk}
\hat{E}_{ll}
\hat{G}_t,
\end{equation}
where $\hat{G}_t$ implements the single-particle rotation defined by $\bm{U}^t$.

Following Ref.~\citenum{von2021quantum}, we assume that $V^{t}_{kl}$ is rank one and positive semidefinite, so that
\begin{equation}
V^{t}_{kl} = W_k^t W_l^t.
\end{equation}

After mapping to Pauli operators, additional one-body contributions arise from normal ordering.
These can be absorbed into the effective one-electron tensor
\begin{equation}
f_{pq}
=
t_{pq}
+
\sum_{r=0}^{n-1}
g_{pqrr}.
\end{equation}

Diagonalizing $\bm{f}$ with eigenvalues $f_k^o$, the Hamiltonian becomes
\begin{equation}
\hat{\mathcal{H}}_{\mathrm{DF}}
=
\beta_{\mathrm{DF}}\mathbb{1}
-
\frac{1}{2}
\sum_{k}
f_k^{o}
\, \hat{G}_{o}^{\dagger}
\left( \hat{Z}_{k} + \hat{Z}_{\bar{k}} \right)
\hat{G}_{o}
+
\frac{1}{8}
\sum_{t=1}^{N_{\mathrm{DF}}}
\hat{G}_{t}^{\dagger}
\left(
\sum_{k}
W_{k}^{t}
\left( \hat{Z}_{k} + \hat{Z}_{\bar{k}} \right)
\right)^{2}
\hat{G}_{t}.
\end{equation}

The one-norm and shift are therefore
\begin{equation}
\lambda_{\mathrm{DF}}
=
\sum_{k} \left| f_k^{o} \right|
+
\frac{1}{4}
\sum_{t=1}^{N_{\mathrm{DF}}}
\left(
\sum_{k}
\left| W_k^{t} \right|
\right)^2,
\end{equation}
and
\begin{equation}
\beta_{\mathrm{DF}}
=
\left| \sum_{k} f^{o}_k
-
\frac{1}{2}
\sum_{t=1}^{N_{\mathrm{DF}}}
\left(
\sum_{k}
W_k^{t}
\right)^2 \right|.
\end{equation}

\subsection{Tensor hypercontraction}
\label{sec:thc}

In the \gls{thc} representation\cite{lee2021even}, the two-electron tensor is approximated as
\begin{equation}
g_{pqrs}
\approx
\sum_{\mu,\nu=1}^{M}
\chi_p^{(\mu)}
\chi_q^{(\mu)}
\,
\zeta_{\mu\nu}
\,
\chi_r^{(\nu)}
\chi_s^{(\nu)},
\end{equation}
where $\zeta_{\mu\nu} = \zeta_{\nu\mu}$ and all tensors are real.
$M=\mathcal{O}(n)$ is the \gls{thc} rank, chosen according to the desired accuracy.
The matrices $\bm{\chi}$ are not orthogonal, but their columns are normalized,
\begin{equation}
\sum_p \chi_p^{(\mu)} \chi_p^{(\mu)} = 1.
\end{equation}

This allows each linear combination $\sum_p \chi_p^{(\mu)} \hat{a}_{p\sigma}$ to be embedded into a unitary transformation acting on fermionic operators, $\hat{\mathcal{U}}^{\dagger}_{\mu}\hat{a}_{0\sigma}\hat{\mathcal{U}}_{\mu}$.
The Hamiltonian can then be written as
\begin{equation}
\hat{\mathcal{H}}
=
\sum_{p,q} t_{pq} \hat{E}_{pq}
+
\frac{1}{2}
\sum_{\mu,\nu}
\zeta_{\mu\nu}\,
\hat{\mathcal{U}}_{\mu}^{\dagger}
\hat{E}_{00}
\hat{\mathcal{U}}_{\mu}
\hat{\mathcal{U}}_{\nu}^{\dagger}
\hat{E}_{00}
\hat{\mathcal{U}}_{\nu}.
\end{equation}

After Jordan--Wigner mapping, one obtains
\begin{equation}
\begin{split}
\hat{\mathcal{H}}_{\mathrm{THC}}
&=
\beta_{\mathrm{THC}} \mathbb{1}
-
\frac{1}{2}
\sum_{k}
f_k^o
\,
\hat{\mathcal{U}}_{o,k}^{\dagger}
\left( \hat{Z}_{k, \alpha} + \hat{Z}_{k,\beta} \right)
\hat{\mathcal{U}}_{o,k}
\\
&\quad
+
\frac{1}{8}
\sum_{\mu,\nu}
\zeta_{\mu\nu}
\,
\hat{\mathcal{U}}_{\mu}^{\dagger}
\left( \hat{Z}_{0, \alpha} + \hat{Z}_{0, \beta} \right)
\hat{\mathcal{U}}_{\mu}
\hat{\mathcal{U}}_{\nu}^{\dagger}
\left( \hat{Z}_{0, \alpha} + \hat{Z}_{0,\beta} \right)
\hat{\mathcal{U}}_{\nu}.
\end{split}
\end{equation}

The corresponding shift and one-norm are
\begin{equation}
\beta_{\mathrm{THC}}
=
\left|\sum_k f^o_k
-
\frac{1}{2}
\sum_{\mu,\nu}
\zeta_{\mu\nu}\right|,
\end{equation}
and
\begin{equation}
\lambda_{\mathrm{THC}}
=
\sum_k |f^o_k|
+
\frac{1}{2}
\sum_{\mu,\nu}
|\zeta_{\mu\nu}|.
\end{equation}

\subsection{Block-invariant symmetry shift (BLISS)}
\label{sec:bliss}

The \gls{bliss} technique\cite{loaiza2023block, patel2025global} modifies the Hamiltonian as
\begin{equation}
\hat{\mathcal{H}}_{\mathrm{BI}}
=
\hat{\mathcal{H}}
-
\alpha_1 \hat{N}
-
\frac{\alpha_2}{2} \hat{N}^2
-
\frac{1}{2} \hat{B}(\hat{N} - \eta),
\end{equation}
where $\hat{N}$ is the particle-number operator, $\eta$ is the number of particles in the symmetry sector of interest and
\begin{equation}
\hat{B}
=
\sum_{\sigma}
\sum_{p,q}
\beta_{pq}
\hat{a}_{p,\sigma}^{\dagger}
\hat{a}_{q,\sigma}
\end{equation}
is a Hermitian one-body operator.

The parameters $\alpha_1$, $\alpha_2$, and $\beta_{pq}$ are optimized to reduce the one-norm of $\hat{\mathcal{H}}_{\mathrm{BI}}$ while preserving the spectrum in the target particle-number sector.

In the $\hat{E}_{pq}$ representation, the BLISS Hamiltonian reads
\begin{equation}
\hat{\mathcal{H}}_{\mathrm{BLISS}} =
\sum_{p,q}
t_{pq}^{(\mathrm{BLISS})}
\hat{E}_{pq}
+
\frac{1}{2}
\sum_{p,q,r,s}
g_{pqrs}^{(\mathrm{BLISS})}
\hat{E}_{pq}
\hat{E}_{rs},
\end{equation}
with
\begin{equation}
t_{pq}^{(\mathrm{BLISS})}
=
t_{pq}
-
\alpha_1 \delta_{pq}
+
\frac{1}{2}
\beta_{pq} \eta,
\end{equation}
and
\begin{equation}
g_{pqrs}^{(\mathrm{BLISS})}
=
g_{pqrs}
-
\alpha_2 \delta_{pq} \delta_{rs}
-
\frac{1}{2}
\left(
\beta_{pq} \delta_{rs}
+
\delta_{pq} \beta_{rs}
\right).
\end{equation}

The THC-BLISS approach\cite{caesura2025faster} then consists of applying the tensor hypercontraction factorization to $t_{pq}^{(\mathrm{BLISS})}$ and $g_{pqrs}^{(\mathrm{BLISS})}$ as described in Sec.~\ref{sec:thc}.

\section{\gls{qksd} Subsampling and Optimal Step Size $\Delta k$}
\label{sec:qksd_subsampling}

Let us consider a physical (shifted) Hamiltonian $\hat{\mathcal{H}}$ and its normalized counterpart $\hat{H} = \hat{\mathcal{H}} / \lambda$, where $\lambda$ denotes the one-norm of $\hat{\mathcal{H}}$.
$\hat{\mathcal{H}}$ eigenvectors are then those of $\hat{H}$ i.e. $\{\ket{\lambda_i}\}_{i=0}^{N-1}$ and its eigenvalues $\{E_n\}_{i=0}^{N-1}$ satisfy the relation $\lambda_n = E_n / \lambda$.
In \gls{qksd}, the overlap matrix elements between Chebyshev–Krylov vectors are
\begin{equation}
\widetilde{S}_{k,k+\Delta k}
:=\bra{\psi_0}T_k(\hat{H})\,T_{k+\Delta k}(\hat{H})\ket{\psi_0}.
\label{eq:exact_S}
\end{equation}

Let us recall the decomposition of $\ket{\psi_0}$ in the eigenbasis of $\hat{H}$ 
\(
\ket{\psi_0}=\sum_n a_n \ket{\lambda_n},
\)
with $a_n\coloneqq \braket{\psi_0|\lambda_n}$ and the corresponding probability distribution $\rho(\tilde{E}) := \sum_n |a_n|^2\, \delta(\tilde{E}-\lambda_n)$.
Since $T_k(\hat{H})\ket{\phi_n} = T_k(\lambda_n)\ket{\lambda_n}$, the overlap matrix elements become
\begin{equation}
    \widetilde{S}_{k,k+\Delta k}
    = \sum_n |a_n|^2\,T_k(\lambda_n)\,T_{k+\Delta k}(\lambda_n).
    \label{eq:S_discrete}
\end{equation}
Eq.~\eqref{eq:S_discrete} may be rewritten as the exact integral
\begin{equation}
\widetilde{S}_{k,k+\Delta k}
= \int_{-1}^{1} T_k(\tilde{E})\,T_{k+\Delta k}(\tilde{E})\,\rho(\tilde{E})\,d\tilde{E}.
\label{eq:S_measure}
\end{equation}

We assume that 

\begin{itemize}
\item $\rho(\tilde{E})$ is supported on a narrow energy window
      $[\tilde{E}_a,\tilde{E}_b]$ centered at $\tilde{E}_*$;
\item the weights $|a_n|^2$ vary slowly across this window,
      so that $\rho(\tilde{E})$ is approximately bounded and smooth;
\end{itemize}

then 
\begin{align}
\widetilde{S}_{k,k+\Delta k}
&= \int_{\tilde{E}_a}^{\tilde{E}_b} T_k(\tilde{E})\,T_{k+\Delta k}(\tilde{E})\,\rho(\tilde{E})\,d\tilde{E} \\
&\approx \rho(\tilde{E}_*) \int_{\tilde{E}_a}^{\tilde{E}_b} T_k(\tilde{E})\,T_{k+\Delta k}(\tilde{E})\,d\tilde{E}\\
&\leq \int_{\tilde{E}_a}^{\tilde{E}_b} T_k(\tilde{E})\,T_{k+\Delta k}(\tilde{E})\,d\tilde{E}.
\label{eq:flat_int}
\end{align}

We now analyze the dependence of two Chebyshev vectors restricted to a given energy window, consider the overlap integral defined in the physical energy space
\begin{equation}
I(\Delta k)
=\int_{E_a}^{E_b} \cos\!\big(k\,\theta(E)\big)\,
                   \cos\!\big((k+\Delta k)\,\theta(E)\big)\, dE,
\label{eq:I_def_energy}
\end{equation}
where
\begin{equation}
\theta(E) = \arccos\!(\frac{E}{\lambda}), \qquad
E = \lambda\cos\theta, \qquad
dE = -\lambda\sin\theta\,d\theta.
\end{equation}
Under this change of variables, Eq.~\eqref{eq:I_def_energy} becomes
\begin{equation}
I(\Delta k)
= \lambda\!\int_{\theta_b}^{\theta_a}
   \sin\theta\,\cos(k\theta)\cos((k+\Delta k)\theta)\,d\theta,
\label{eq:I_theta}
\end{equation}
with $\theta_i=\arccos(E_i/\lambda)$.

Using the trigonometric identity $\cos A\cos B = \tfrac12[\cos(A-B)+\cos(A+B)]$, the integrand can be rewritten as
\begin{equation}
\cos(k\theta)\cos((k+\Delta k)\theta)
= \tfrac12\!\left[\cos(\Delta k\,\theta)
                 + \cos((2k+\Delta k)\theta)\right],
\end{equation}
yielding
\begin{equation}
I(\Delta k)
= \frac{\lambda}{2}\!
  \int_{\theta_b}^{\theta_a}\!\!\sin\theta\,
   \cos(\Delta k\,\theta)\,d\theta
+ \frac{\lambda}{2}\!
  \int_{\theta_b}^{\theta_a}\!\!\sin\theta\,
   \cos((2k+\Delta k)\theta)\,d\theta.
\label{eq:I_split}
\end{equation}
The first term varies slowly with $\theta$, its frequency set by $\Delta k$, while the second term oscillates rapidly with frequency $2k+\Delta k$.
In the Krylov or Chebyshev context one typically has $k\gg1$ and $\Delta k\ll 2k$, so the second term integrates to a much smaller value.

Physically, $\cos(k\theta)\cos((k+\Delta k)\theta)$ represents the interference of two oscillations: the term $\cos(\Delta k\,\theta)$ is the \emph{slow envelope} of the interference, while $\cos((2k+\Delta k)\theta)$ is a \emph{high--frequency carrier} that averages out under integration.
Hence we keep only the envelope term,
\begin{equation}
I(\Delta k)\;\approx\;
\frac{\lambda}{2}\int_{\theta_b}^{\theta_a}
  \sin\theta\,\cos(\Delta k\,\theta)\,d\theta.
\label{eq:I_slow}
\end{equation}

If the window $[E_a,E_b]$ (or equivalently $[\theta_b,\theta_a]$) is narrow, $\sin\theta$ can be replaced by its value at the midpoint $\theta_\ast$, giving
\begin{equation}
I(\Delta k)
\simeq
\frac{\lambda}{2}\sin\theta_\ast
\int_{\theta_b}^{\theta_a}\cos(\Delta k\,\theta)\,d\theta
= \frac{\lambda}{2}\sin\theta_\ast
  \frac{\sin(\Delta k\,\theta_a)-\sin(\Delta k\,\theta_b)}{\Delta k}.
\label{eq:I_small_window}
\end{equation}

The difference of sines in Eq.~\eqref{eq:I_small_window} can be written as
\[
\sin(\Delta k\,\theta_a) - \sin(\Delta k\,\theta_b)
= 2\cos\!\left(\frac{\Delta k(\theta_a+\theta_b)}{2}\right)
  \sin\!\left(\frac{\Delta k(\theta_a-\theta_b)}{2}\right).
\]
Substituting this into Eq.~\eqref{eq:I_small_window} gives
\begin{equation}
I(\Delta k)
\simeq
\lambda\sin\theta_\ast
\frac{\cos\!\left(\tfrac{\Delta k(\theta_a+\theta_b)}{2}\right)
      \sin\!\left(\tfrac{\Delta k(\theta_a-\theta_b)}{2}\right)}
{\Delta k}.
\label{eq:I_simplified}
\end{equation}
The overlap $I(\Delta k)$ vanishes when $\Delta k(\theta_a+\theta_b)/2 = \frac{\pi}{2}$ or $\Delta k(\theta_a-\theta_b)/2 = \pi$.
In addition, the overlap decays with the magnitude 
\begin{equation}
    |I(\Delta k)| \leq \frac{\lambda \sin\theta_*}{\Delta k}.
    \label{eq:ovlp_mag}
\end{equation}

Hence, the choice 
\begin{equation}
    \Delta k = \lambda \sin \theta_* = \lambda \sqrt{1-\tilde{E}_*^2}
\end{equation}
always ensures that the overlap between two Chebyshev polynomials whose orders differ by $\Delta k$ is small. 
The $\sqrt{1-\tilde{E}_*^2}$ factor reduces the step size if the relevant eigenstates lie close to the spectrum edges. 

\section{Spectral Analysis of Krylov overlap matrix}
\label{sec:appendix_seigs_growth}

In this appendix we analyze how the dominant eigenvalues and eigenvectors of the Krylov overlap matrix evolve as the maximal Chebyshev degree $K$ increases. This provides a geometric explanation for the behavior observed in Figure~\ref{fig:noisy_qksd_fe2s2} and clarifies why, for sufficiently large $K$, only a small number of overlap eigenvalues need to be retained.

Throughout this section, we consider the Chebyshev Krylov space
\begin{equation}
\mathcal{K} = \mathrm{span}\{ T_k(\hat H)\ket{\psi_0} \}_{k=0}^{K-1},
\end{equation}
where $\hat H$ is the shifted and normalized Hamiltonian satisfying $\|\hat H\|\le 1$.

\subsection{Spectral representation of the overlap matrix}

Let the initial state be expanded in the eigenbasis of $\hat H$,
\begin{equation}
\ket{\psi_0} = \sum_{r=0}^{R-1} \sqrt{p_r}\,\ket{\lambda_r},
\qquad
p_r = |\langle \lambda_r|\psi_0\rangle|^2.
\end{equation}
The Chebyshev moments entering the \gls{qksd} matrices are
\begin{equation}
\langle T_k(\hat H) \rangle
=
\sum_{r=0}^{R-1} p_r\, T_k(\lambda_r).
\end{equation}
The overlap matrix elements are
\begin{equation}
\widetilde{S}_{kj}
=
\langle \psi_0| T_k(\hat H) T_j(\hat H) |\psi_0\rangle
=
\sum_{r=0}^{R-1} p_r\, T_k(\lambda_r) T_j(\lambda_r).
\label{eq:Skj_spectral}
\end{equation}
Defining the Chebyshev feature vectors
\begin{equation}
v_r^{(K)} :=
\begin{pmatrix}
T_0(\lambda_r) \\
T_1(\lambda_r) \\
\vdots \\
T_{K-1}(\lambda_r)
\end{pmatrix}
\in \mathbb{R}^{K},
\end{equation}
the overlap matrix admits the compact representation
\begin{equation}
\widetilde{S} = \sum_{r=0}^{R-1} p_r\, v_r^{(K)} \left( v_r^{(K)} \right)^T.
\label{eq:S_gram}
\end{equation}

Eq.~\eqref{eq:S_gram} shows that $\widetilde{S}$ is a weighted Gram matrix of Chebyshev feature vectors evaluated on the Hamiltonian spectrum. Its rank is therefore at most $R$, independently of $K$.

\subsection{Growth of dominant overlap eigenvalues with $K$}

The Gram representation in Eq.~\eqref{eq:S_gram} provides a direct geometric interpretation of the spectrum of the overlap matrix. Each term
\begin{equation}
p_r\, v_r^{(K)} (v_r^{(K)})^T
\end{equation}
is a rank-one matrix whose only nonzero eigenvalue equals
\begin{equation}
p_r \| v_r^{(K)} \|^2,
\end{equation}
with eigenvector proportional to $v_r^{(K)}$, since
\begin{equation}
p_r v_r^{(K)} (v_r^{(K)})^T v_r^{(K)}
=
p_r \|v_r^{(K)}\|^2 v_r^{(K)}.
\end{equation}

The full overlap matrix is a sum of such rank-one contributions. Although the vectors $v_r^{(K)}$ are not mutually orthogonal, their inner products remain bounded as $K$ increases, while their squared norms grow linearly with $K$. This separation of scales controls the asymptotic behavior of the eigenvalues.

To make this explicit, write $\lambda_r = \cos \theta_r$ with $\theta_r \in [0,\pi]$. Then
\begin{equation}
T_k(\lambda_r) = \cos(k\theta_r),
\end{equation}
and the inner product between two feature vectors becomes
\begin{equation}
v_r^{(K)\,T} v_s^{(K)}
=
\sum_{k=0}^{K-1} \cos(k\theta_r)\cos(k\theta_s).
\label{eq:inner_product}
\end{equation}
Using $\cos a \cos b = \tfrac12[\cos(a-b)+\cos(a+b)]$, we obtain
\begin{equation}
v_r^{(K)\,T} v_s^{(K)}
=
\frac12 \sum_{k=0}^{K-1} \cos(k(\theta_r-\theta_s))
+
\frac12 \sum_{k=0}^{K-1} \cos(k(\theta_r+\theta_s)).
\end{equation}
Each sum admits the closed form
\begin{equation}
\sum_{k=0}^{K-1} \cos(k\phi)
=
\frac{\sin(K\phi/2)}{\sin(\phi/2)}
\cos\!\left(\frac{(K-1)\phi}{2}\right),
\end{equation}
which remains bounded as $K\to\infty$ whenever $\phi \neq 0$. Therefore, for $r\neq s$,
\begin{equation}
v_r^{(K)\,T} v_s^{(K)} = \mathcal{O}(1).
\label{eq:offdiag}
\end{equation}
In contrast, the diagonal term satisfies
\begin{equation}
\| v_r^{(K)} \|^2
=
\sum_{k=0}^{K-1} \cos^2(k\theta_r)
=
\frac{K}{2}
+
\mathcal{O}(1),
\label{eq:diag}
\end{equation}
since $\cos^2 x = \tfrac12(1+\cos 2x)$.

Equations~\eqref{eq:offdiag} and \eqref{eq:diag} show that diagonal contributions grow linearly with $K$, while cross terms remain bounded. Consequently, in the large-$K$ regime the rank-one contributions become asymptotically orthogonal in feature space, and the dominant eigenvalues of $\widetilde{S}$ inherit the leading scaling
\begin{equation}
s_r
=
p_r \| v_r^{(K)} \|^2
+
\mathcal{O}(1)
\sim
\frac{p_r}{2}\,K.
\label{eq:sr_scaling}
\end{equation}

The dominant overlap eigenvalues therefore grow approximately linearly with the maximal Chebyshev order $K$, with slopes proportional to the spectral weights $p_r$ of the initial state. This linear growth explains the monotonic increase of the leading eigenvalues observed numerically in Figure~\ref{fig:noisy_qksd_fe2s2}(a).

\subsection{Emergence of spectral separation}

Eq.~\eqref{eq:sr_scaling} also clarifies the emergence of spectral gaps in the overlap matrix. For two components $r$ and $t$,
\begin{equation}
s_r - s_t
\sim
\frac{K}{2}\,(p_r - p_t).
\end{equation}

As long as $p_0 > p_1$, the separation between the largest two eigenvalues increases linearly with $K$. The relative ordering of overlap eigenvalues is therefore determined entirely by the ordering of spectral weights in the initial state.

Consequently:
\begin{itemize}
\item The leading overlap eigenvector becomes increasingly aligned with the Chebyshev feature vector of the ground state.
\item The gap between the first few eigenvalues widens with $K$.
\item The smallest retained eigenvalue grows proportionally to $K$, lowering the precision requirement needed to resolve it from statistical noise.
\end{itemize}

This behavior provides a geometric explanation for the sharp decrease in required shot counts at larger $K$ reported in Figure~\ref{fig:noisy_qksd_fe2s2}(c).

\subsection{Non-monotonic behavior of the regularized energy error}

While the leading overlap eigenvalues grow approximately linearly with $K$, as shown above, the corresponding ground-state energy error after regularization (Figure~\ref{fig:noisy_qksd_fe2s2}(b)) is not strictly monotonic in $K$. This behavior follows from the fact that the regularized QKSD estimator is not variational and does not act on a nested sequence of subspaces.

Indeed, although the Krylov spaces satisfy
\begin{equation}
\mathcal{K}_K \subset \mathcal{K}_{K+1},
\end{equation}
the regularized procedure retains only the $m$ dominant eigenvectors of the overlap matrix. Denoting by $P_m(K)$ the projector onto the subspace spanned by these $m$ eigenvectors, the effective subspace used for the generalized eigenvalue problem is
\begin{equation}
\widetilde{\mathcal{K}}_K = P_m(K)\,\mathcal{K}_K.
\end{equation}

Since the dominant eigenvectors of $S(K)$ rotate as $K$ increases, the projectors $P_m(K)$ do not form a nested sequence. Consequently,
\begin{equation}
\widetilde{\mathcal{K}}_K \not\subset \widetilde{\mathcal{K}}_{K+1}
\end{equation}
in general, and the associated Ritz estimates of the ground-state energy are not guaranteed to improve monotonically.

Thus, while increasing $K$ systematically improves the conditioning of the overlap matrix and enhances spectral separation, the combination of regularization and truncation induces a $K$-dependent effective subspace whose geometry may fluctuate slightly before stabilizing. This explains the non-monotonic features observed in Figure~\ref{fig:noisy_qksd_fe2s2}(b).

\section{Additional Sampling-Noise Analysis for Fe$_4$S$_4$, Co(salophen) and naphthalene}
\label{sec:appendix_qksd_noise}

Here we report additional numerical results for Fe$_4$S$_4$ Co(salophen) and naphthalene, complementing the detailed analysis presented in Section~\ref{sec:shot_noise_numerics} for Fe$_2$S$_2$. The goal is to verify that the observed interplay between overlap eigenvalues, regularization, and sampling cost is not system-specific but represents a generic feature of \gls{qksd}.

\begin{figure}[h!]
    \centering
    \begin{subfigure}{0.48\textwidth}
        \centering
        \includegraphics[width=\linewidth]{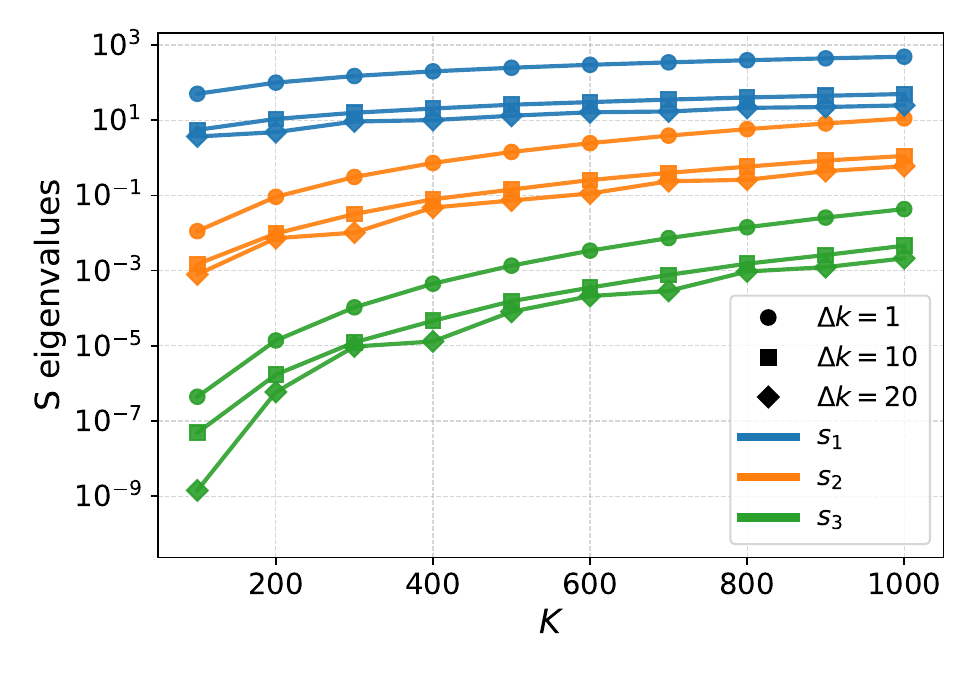}
        \caption{Fe$_4$S$_4$}
    \end{subfigure}
    \hfill
    \begin{subfigure}{0.48\textwidth}
        \centering
        \includegraphics[width=\linewidth]{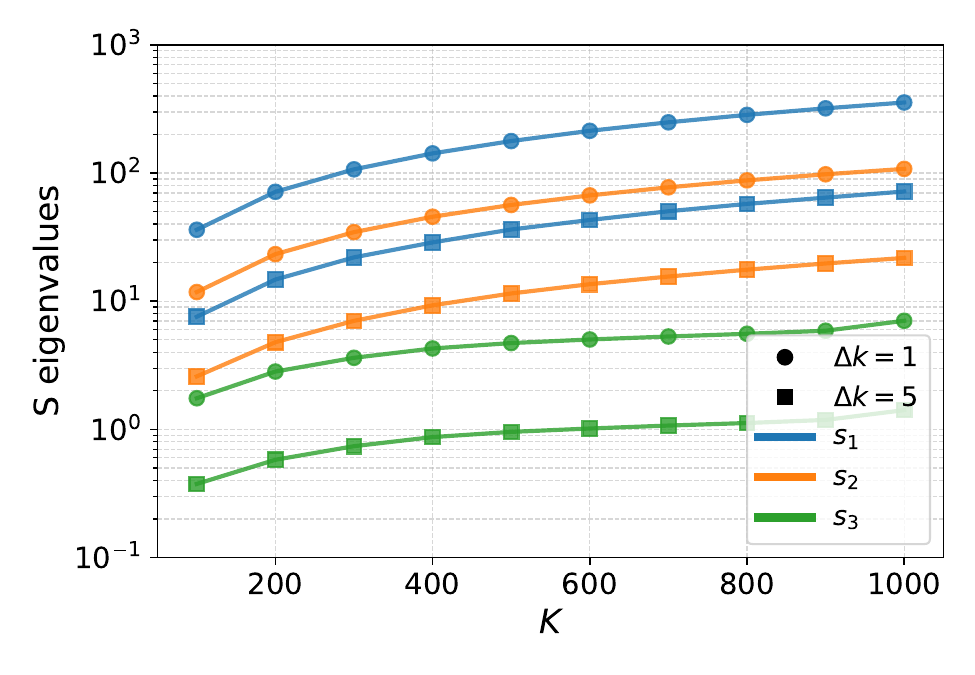}
        \caption{Naphthalene}
    \end{subfigure}
    \medskip
    \begin{subfigure}{0.48\textwidth}
        \centering
        \includegraphics[width=\linewidth]{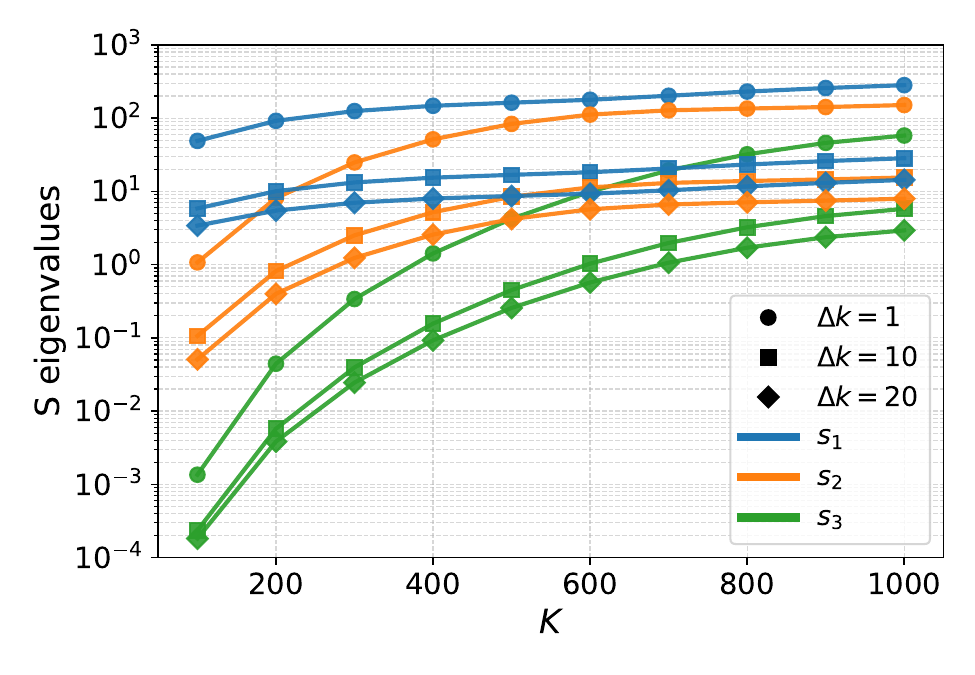}
        \caption{Co(salophen)}
    \end{subfigure}
    \caption{
    Three largest eigenvalues of the overlap matrix $\tilde{S}$ as a function of the maximal Chebyshev order $K$ for different polynomial step sizes $\Delta k$. For all systems, increasing $\Delta k$ reduces the magnitude of the retained overlap eigenvalues due to principal subspace interlacing.
    }
    \label{fig:appendix_overlap}
\end{figure}

Figure~\ref{fig:appendix_overlap} shows the three largest eigenvalues of the overlap matrix $\tilde{S}$ as a function of the maximal Chebyshev order $K$ for different polynomial step sizes $\Delta k$. As for Fe$_2$S$_2$, the leading eigenvalues increase monotonically with $K$, reflecting the growing linear independence of the Krylov vectors. 

For fixed $K$, we again observe the systematic ordering
$
s_i^{\Delta k = 1}
>
s_i^{\Delta k = 10}
>
s_i^{\Delta k = 20},
$
which follows from the Poincar\'e separation theorem: restricting to a principal subspace via coarse polynomial sampling necessarily reduces the magnitude of the overlap eigenvalues.

\begin{figure}[h!]
    \centering
    \begin{subfigure}{0.48\textwidth}
        \centering
        \includegraphics[width=\linewidth]{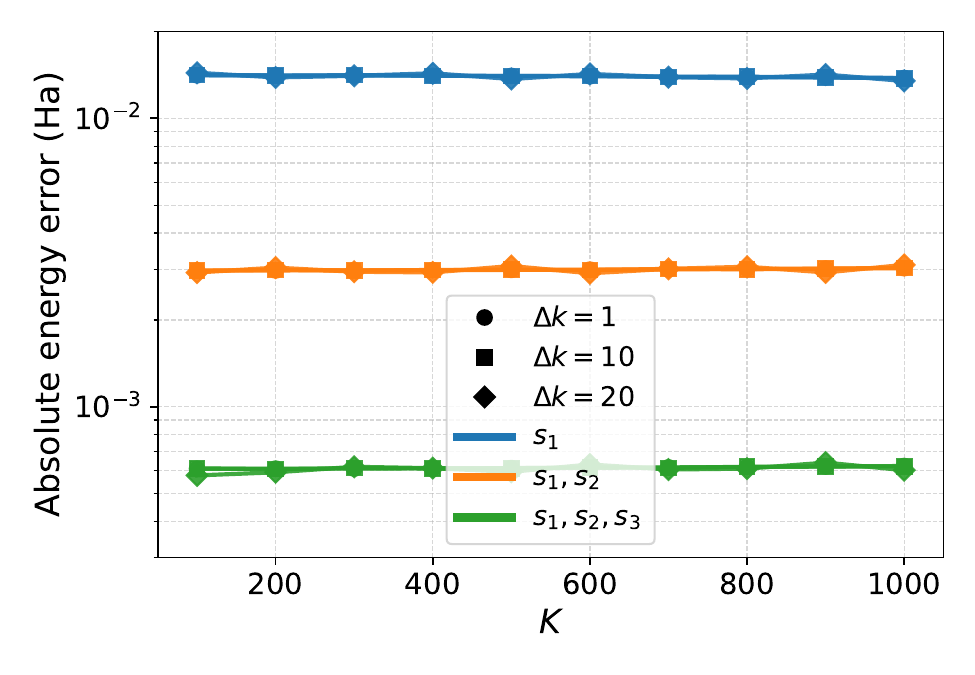}
        \caption{Fe$_4$S$_4$}
    \end{subfigure}
    \hfill
    \begin{subfigure}{0.48\textwidth}
        \centering
        \includegraphics[width=\linewidth]{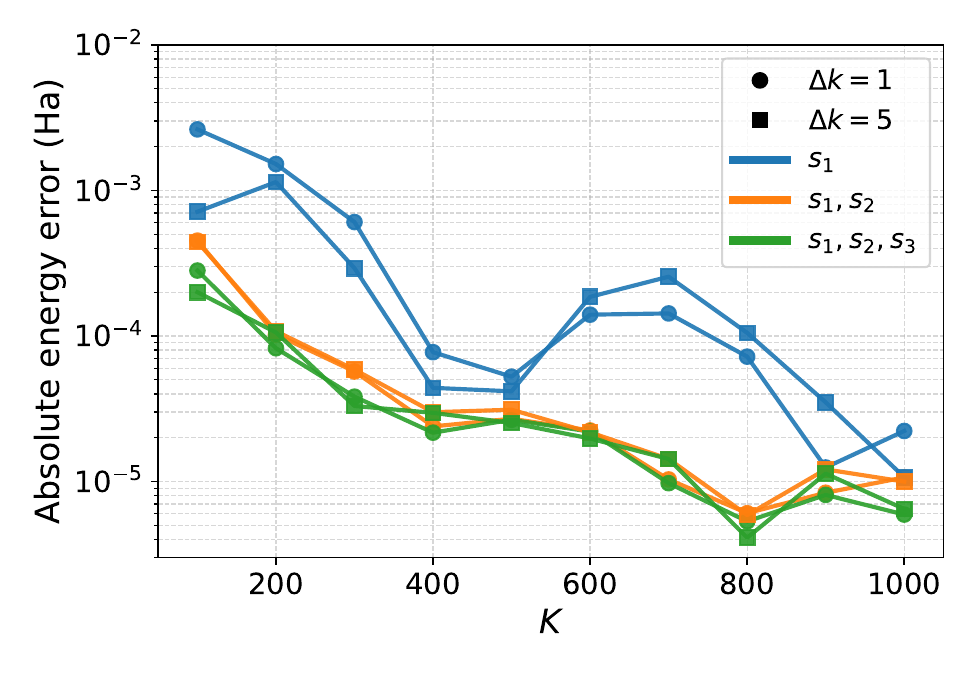}
        \caption{Naphthalene}
    \end{subfigure}
    \medskip
    \begin{subfigure}{0.48\textwidth}
        \centering
        \includegraphics[width=\linewidth]{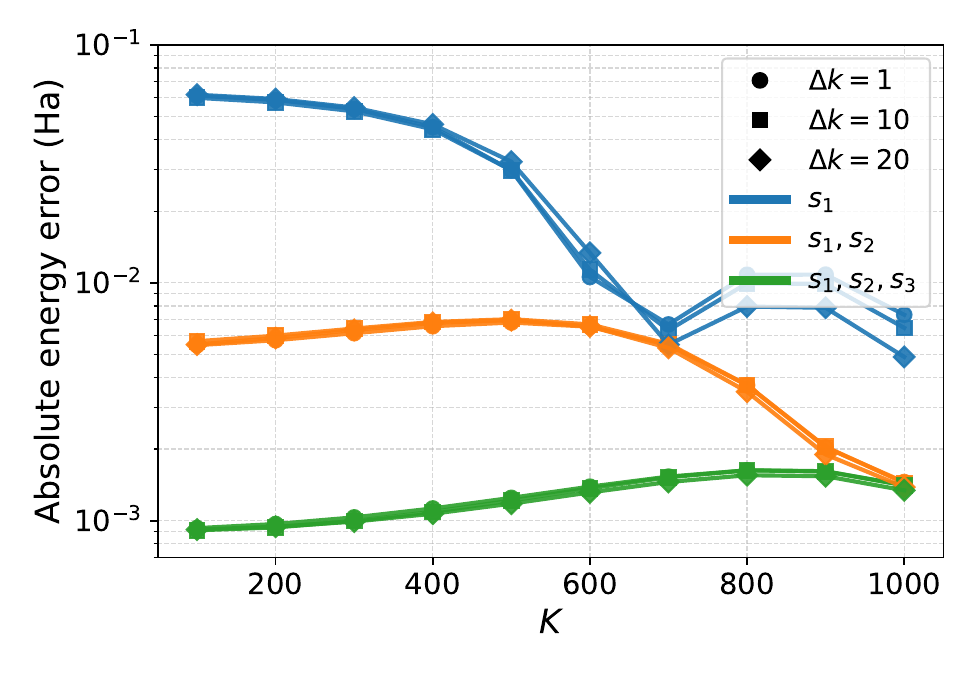}
        \caption{Co(salophen)}
    \end{subfigure}
    \caption{
    Absolute ground-state energy error induced by regularization of the generalized eigenvalue problem when retaining the largest one, two, or three eigenvalues of $\tilde{S}$. The systematic bias decreases with increasing $K$ and is qualitatively similar across systems.
    }
    \label{fig:appendix_bias}
\end{figure}

Figure~\ref{fig:appendix_bias} reports the absolute ground-state energy error obtained when solving the generalized eigenvalue problem while retaining only the largest one, two, or three eigenvalues of $\tilde{S}$.
As in the Fe$_2$S$_2$ case, truncating the Krylov subspace introduces a systematic bias which we observe to decreases non-monotonically with $K$. Retaining three eigenvalues is generally sufficient to reach chemical accuracy for the systems considered here.

\begin{figure}[h!]
    \centering
    \begin{subfigure}{0.48\textwidth}
        \centering
        \includegraphics[width=\linewidth]{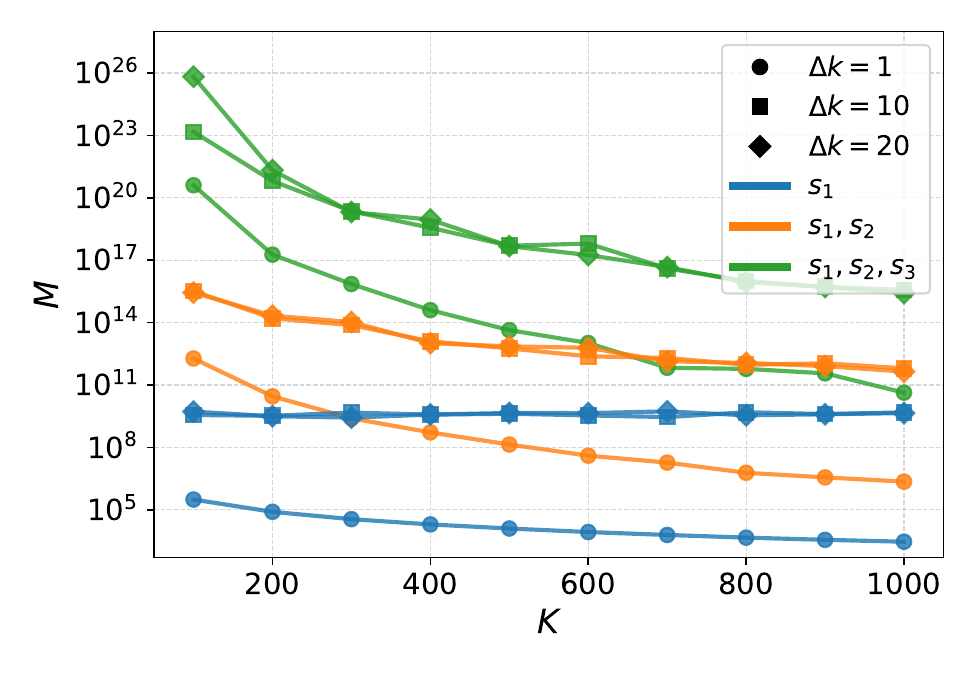}
        \caption{Fe$_4$S$_4$}
    \end{subfigure}
    \hfill
    \begin{subfigure}{0.48\textwidth}
        \centering
        \includegraphics[width=\linewidth]{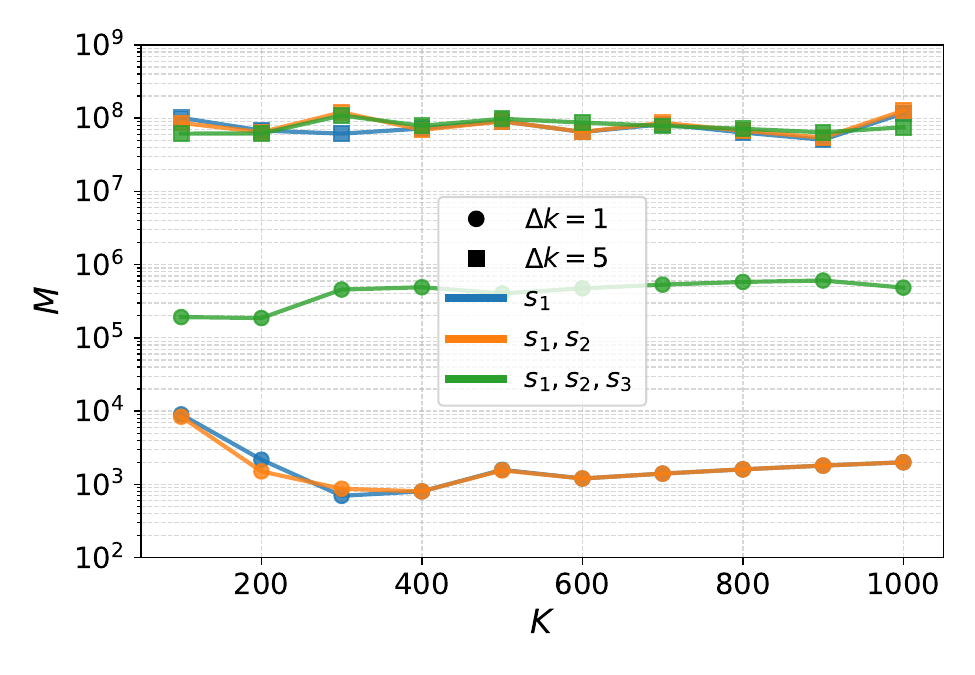}
        \caption{Naphthalene}
    \end{subfigure}
    \medskip
    \begin{subfigure}{0.48\textwidth}
        \centering
        \includegraphics[width=\linewidth]{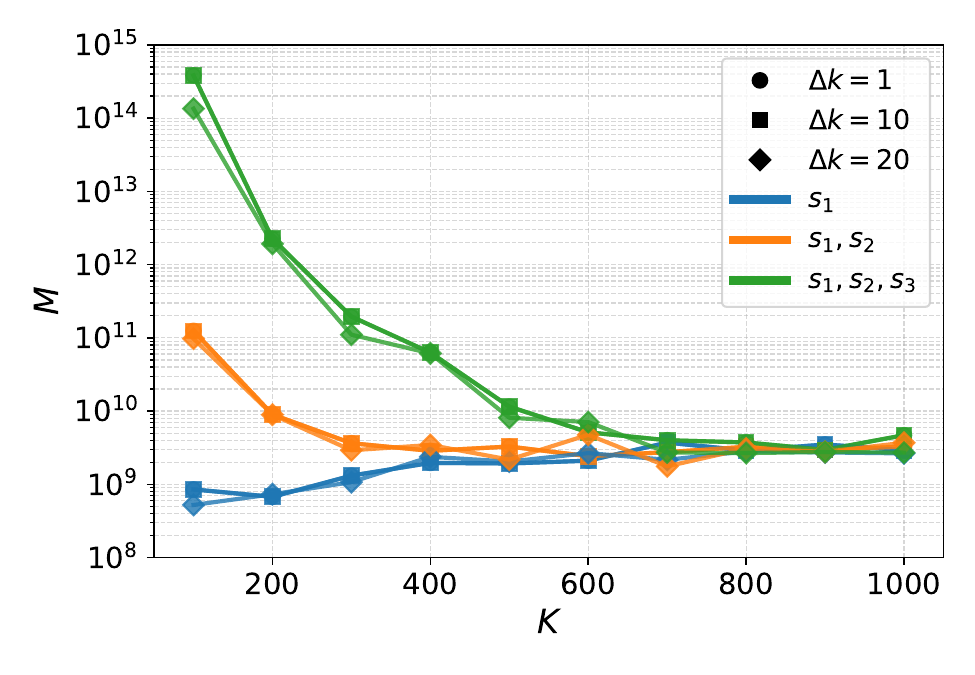}
        \caption{Co(salophen)}
    \end{subfigure}
    \caption{
    Total number of shots required to reach 1 mHa accuracy, on average over 100 trials, relative to the noiseless regularized reference using gradient-based shot allocation. Although increasing $\Delta k$ reduces the number of measured Chebyshev moments, the reduced magnitude of the overlap eigenvalues leads to a higher overall sampling cost.
    }
    \label{fig:appendix_sampling}
\end{figure}

Finally, Figure~\ref{fig:appendix_sampling} shows the total number of shots required to achieve 1 mHa accuracy, on average over 100 trials, with respect to the noiseless regularized reference. The shot budget is determined using the same gradient-based allocation strategy described in Section~\ref{sec:shot_noise_numerics}.

Consistent with the Fe$_2$S$_2$ results, increasing $\Delta k$ reduces the number of measured Chebyshev moments but simultaneously decreases the magnitude of the retained overlap eigenvalues. Because resolving smaller eigenvalues requires suppressing statistical fluctuations below a lower threshold, the total shot count increases. This confirms that subsampling in polynomial order does not reduce the overall sampling cost once shot noise is taken into account.

Overall, these results demonstrate that the trade-off between Krylov dimension, overlap eigenvalue magnitude, and sampling complexity is robust across different molecular systems.

The achieved energy accuracy at the largest subspace dimension in each of the cases, in the noise-free scenario i.e. bias, is reported in Table~\ref{tab:qksd_data}.

\begin{table}[]
    \centering
    \begin{tabular}{l|r|r|r|r}
         & $\mathrm{Fe_4S_4} (54, 36)$ & $\mathrm{Fe_2S_2} (30, 20)$ & Co(salophen) (26, 27) & Naphthalene (10, 10) \\
         \hline
     $\Delta E_0$ & 0.014105 &  0.038936 & 0.061914 & 0.093205  \\    
     \hline 
     $s_1, s_2$ & 0.003480 & 0.000442 & 0.000424 & 0.000011 \\
     \hline 
     $s_1, s_2, s_3$ & 0.000961 & 0.000116 & 0.000425 & 0.000002 \\
     \hline 
     $\lambda_{\mathrm{THC-BLISS}}$ & 63.355 & 24.390 & 28.132 & 5.492 \\
     \hline
     $\beta_{\mathrm{THC-BLISS}}$ & 336.353 & 118.880 & 2402.159 & 382.282 \\
     \hline
     $p_0$ & 0.5 & 0.5 & 0.5 & 0.710886   \\
    \end{tabular}
    \caption{
    Energy difference between the Initial state and the ground state energy $\Delta E_0$ in Hartree, systematic (shot noise free) \gls{qksd} error in Hartree when truncating the generalized eigenvalue problem at two ($s_1, s_2$) or three ($s_1, s_2, s_3$) eigenvalues of the $\widetilde{S}$ matrix respectively at the largest K (in Figure~\ref{fig:fig1}) for the different active space Hamiltonians considered in this work as well as the THC-BLISS norm $\lambda_{\mathrm{THC-BLISS}}$, shift $\beta_{\mathrm{THC-BLISS}}$, and initial state overlap $p_0$ (for details on the initial states see Section~\ref{sec:molecules}).
    We used the active space Hamiltonians from Ref.~\citenum{li2017spin} for the iron sulfur complexes and from Ref.~\citenum{Dutkiewicz_2025} for Co(salophen) and Naphthalene.
    }
    \label{tab:qksd_data}
\end{table}
\end{appendices}

\clearpage
\bibliography{main}

\end{document}